\documentstyle[PASJadd,epsbox,12pt]{PASJ95}
\draft
\newcommand{\vol}{}
\newcommand{\no}{}
\newcommand{\lett}{}
\newcommand{\spage}{}
\newcommand{\rdate}{}
\newcommand{\adate}{}

\begin{document}

\title{When the Hubble Sequence Appeared ?: Morphology, Color, and
Number-Density Evolution of the Galaxies in the Hubble Deep Field North.}

\author{ 
Masaru {\sc Kajisawa}\\
and \\
Toru {\sc Yamada} \\
{\it National Astronomical Observatory, 2-21-1,
Osawa, Mitaka, Tokyo 181-8588} \\
{\it E-mail(MK): kajisawa@optik.mtk.nao.ac.jp}
\\[6pt]
} 

%

\abst{Using the HST WFPC2/NICMOS archival data of the Hubble Deep Field North,
we construct the nearly complete sample of the M$_{V}<-20$ ($\sim$L$^* +1$)
galaxies to $z=2$, and investigate when the Hubble sequence appeared, namely,
the evolution of the morphology, colors, and the comoving number density of the
sample. Even if taking into account of the uncertainty of the photometric
redshift technique, the number density of relatively bright bulge-dominated
galaxies in the HDF-N decrease significantly at $z>1$, and their rest-frame
$U-V$ color distribution is wide-spread over $0.5<z<2$. 
On the other hand, while
the number density of both disk-dominated and irregular galaxies does not show
significant change at $z<2$, their distribution of the rest-frame $U-V$ color
alters at $z\sim1.5$: there is no relatively red (rest $U-V\gtsim0.3$) galaxies
at $z>1.5$, while the significant fraction of these red disk-dominated or
irregular galaxies exists at $z<1.5$. 
These results suggest that the significant
evolution of the Hubble sequence which is seen 
in the present Universe occurs at $1<z<2$.
}

\kword{Galaxies: field --- Galaxies: evolution --- Galaxies:formation}

\maketitle

\setlength{\baselineskip}{18pt}
\section{Introduction}

   Since Hubble (1926; 1936), galaxy properties are known to be broadly
classified along the sequence of their morphology, namely, bulge-to-disk
luminosity ratio and the tightness of the spiral arm structure, 
from ellipticals
(E) to irregular (Irr) galaxies. The classification has been repeatedly tested
and expanded (e.g., de Vaucouleurs 1959; Sandage 1975; Buta 1992 a,b; Roberts
and Haynes 1994; see Abraham 1999 for the recent review).
  Along the Hubble sequence, some physical properties of the galaxies changes.
The cold gas to luminous mass ratio changes systematically from Scd to E type
(Roberts and Haynes 1994), and thus the sequence must represent the different
evolutionary history of the galaxies.
 Consequently, the colors and spectral properties also change along the Hubble
sequence. In fact, local galaxies are known to lie on a sequence in $U-V$ and
$B-V$  two color diagram (Huchra 1977; Kennicutt 1983). In other words, galaxy
SED can be characterized by the rest-frame $U-V$ color since this color is
sensitive to the existence/absence of the largest features in galaxy continuum
spectra at near-UV to optical wavelength, namely, 4000 \AA\  
break or the Balmer
discontinuity that represent the average age and/or metallicity of the stellar
population in the galaxies.

 Thanks to the recent developments of technology, photometric properties of
galaxies can be traced toward very high redshift and the evolution of 
the Hubble
sequence can be directly and empirically tested. It is very exciting to observe
when and how the Hubble sequence appeared as we see in the local universe.

 The combinations of redshift survey and high resolution imaging with Hubble
Space Telescope (HST) have revealed that the large, normal galaxies such as
those represented in the Hubble sequence scheme show little or 
mild evolution to $z\sim1$ (e.g., Schade et al. 1996; 
Brinchmann et al. 1998; Lilly et al. 1998).
  On the other hand, the large sample of star-forming galaxies at $z>3$ was
discovered with `drop-out' technique (e.g., Steidel et al. 1996; 1999) and these
Lyman Break Galaxies are found to have the morphology that is clearly different
from Hubble sequence not only in the rest-frame UV wavelength but also in the
rest-frame optical region (Dickinson 2000a).

  From these previous studies, therefore, the intermediate redshift range at $1
< z < 2$ is supposed to be very important era for the formation of the Hubble
sequence.

  However, there are some difficulties in studying a well defined sample of
galaxies at $1 < z < 2$.  Firstly, the optical flux-limited selection of
galaxies is biased to detect more blue star-forming galaxies at $z \gtsim 1.3$
since it samples the rest-frame UV-luminous objects in the redshift range. In
order to study the evolution of Hubble sequence, the sample should contains
quiescent galaxies as the local early-type objects. Sample selection in NIR
wavelength is essential to construct a nearly mass-limited sample to study the
evolution of colors, morphologies, and the number density of the galaxies in
certain mass range. Limited area of the infrared detectors and the high
background sky noise in the ground-base observations, however, 
generally prevent
us from building an ideal sample of galaxies in the redshift range with fairly
large volume and depth.

  It is also difficult to identify the redshift of galaxies in this range by
optical spectroscopy since the major emission lines 
and continuum breaks get out
of optical wavelength. The depth and multiplicity in NIR spectroscopy is still
much limited compared to the optical ones. Alternative way of determining the
redshift of galaxies at $1<z<2$ is that by photometric redshift technique with
multi-band  optical-NIR photometric data.

 Further, galaxy morphology appears differently when seen in the different
wavelength. At high redshift, the morphological K-correction (e.g.,O'connell
1997) seems to be significant in the optical images, particularly for disk
galaxies. NIR high-resolution image that can allow the morphological
classification for galaxies at $z > 1$ is needed to overcome this problem.

  In this paper, in order to understand the formation and evolution of the
Hubble sequence, we construct the nearly complete sample of the $M_{V} < -20$
($\sim$L$^* +1$) galaxies in the HDF-N to $z=2$  using the archival HST/NICMOS
NIR
data in addition to the optical WFPC2 data. We investigate the changes of the
distributions of galaxy morphology, luminosity and colors since $z=2$ with this
sample.
  Thanks to the deep NICMOS images, we can sample galaxies at $1<z<2$ to $M_V
< -20$ with very high completeness, and the high resolution of these images 
also enable us to examine the galaxy morphology in the same rest-frame 
wavelength at each redshift.

  It is known that galaxy morphological fraction changes with the absolute
magnitude range (Bingelli, Sandage, and Tammann 1988; Marzke et al. 1994; van
den Bergh 1997).  Marzke et al. obtained the fitted $L_*$ values of the
luminosity function for each type of the galaxies as $-20.7$ for E and
$\approx -20.2-3$ for S0 to Sm-Im ($H_0=50$ km s$^{-1}$ Mpc$^{-1}$). According
to the van den Bergh (1997) who reviewed the Revised Shapley Ames Catalog for
Bright Galaxies (RSA2, Sandage and Tammann 1987), more than 90$\%$ of the
galaxies with E-Sb and 80$\%$ of the galaxies with Sc in the RSA2 catalog have
absolute magnitude of M$_{V} < -20$ 
though only a 20\% of the
galaxies later than Scd in RSA2 are found in the luminosity range. Although
there is some uncertainty for the luminosity function of the galaxies 
later than
Scd, the limit of $M_{V} < -20$, as the baseline, provides us to recover the
Hubble sequence from E to at least Sc type {\it in the local universe}.

 On the other hand, we also suffer the small volume effect in the current
analysis since the HDF-N covers at most $\sim$ 4 arcmin$^2$. It should be
reminded that the results obtained from such a small field could be easily
affected by large-scale structure. However, the WFPC2 $+$ NICMOS data of the
HDF-N are the wide wavelength coverage, highest resolution, deepest images at
the present, and provide us the nearly the first opportunity to obtain the
implication for the understanding of the formation of the Hubble sequence. Note
that the region also samples the comoving depth of 1.9 Gpc 
between $1<z<2$ although only $\sim$10$^{4}$ Mpc$^{-3}$ 
in comoving volume ($H_0=70$
km s$^{-1}$ Mpc$^{-1}$, $\Omega_{\rm 0} = 0.3$, $\Omega_{\Lambda} = 0.7$).

  Dickinson (2000b) originally studied the same HST WFPC2/NICMOS data of the
HDF-N, and investigated the morphology, luminosity, color, number density of
relatively bright $z<2$ galaxies with similar luminosity range to those studied
here.  They found the similarity between the rest UV and optical morphology of
irregular galaxies, the existence of the large, ordinary spiral galaxies to
$z\sim1.2$ and the red giant ellipticals to $z\sim1.8$, the presence of bluer
early-type galaxies at $0.5<z<1.4$, and the deficit of bright galaxies at $z>1$
in all morphological types.

For the morphologically-selected galaxies, Franceschini et al.(1998)
and Rodighiero et al.(2000) also investigated the redshift distribution of the
galaxies in the HDF-N  and found the deficit of bright objects at $z\gtsim1.3$
in each morphology.

 In this paper, we also tested these results by our own morphological
classification, color analysis in the different wavelength, taking into the
various uncertainties in the sample selection due to the cosmology, photometric
redshift procedure, and band-shifting effect.

 In section 2, we describe the data reduction, and the selection procedure of
our sample.  We investigate the evolution of number density and color
distribution of these galaxies, and check these results, especially about the
uncertainty of the photometric redshift estimate in section 3. The discussion
about these results is presented in section 4,  and we summarize our conclusion
in the 
final section. We use AB magnitude system (Oke 1974) for HST filter
bands, and refer F300W, F450W, F606W, F814W, F110W, F160W bands as $U_{300}$,
$B_{450}$, $V_{606}$, $I_{814}$, $J_{110}$, $H_{160}$, 
respectively, through the paper.

\section{Data Reduction \& Sample Selection}

\subsection{HST WFPC2/NICMOS data}

  The HDF-N is observed with HST NICMOS Camera 3 between UT 1998 June 13 and
June 23 (PI:M. Dickinson; PID 7817). The complete HDF-N was mosaiced with eight
sub fields in $J_{110}$ and $H_{160}$-bands. We analyzed the calibrated data of
these observations downloaded from the archival site of the Space Telescope
Science Institute. Each sub field was observed through the nine dithered
pointings, with a net exposure of 12600 sec in each band, except for a few
pointings that cannot be used due to guidance error of HST (Dickinson et al.
2000).   We combine these data into a single mosaiced image registered to WFPC2
image of HDF-N, using the ``drizzling'' method with $IRAF$ $DITHER$ package. The
FWHM of the final drizzled image is about 0.22 arcsec.

 For optical data, we analyzed the public ``version 2'' $U_{300}$, $B_{450}$,
$V_{606}$, $I_{814}$-band WFPC2 images of HDF-N produced by the STScI team. The
optical images were convolved with Gaussian kernel to match the NICMOS point
spread function.

\subsection{Source Detection \& Photometry}

 At first, we performed source detection in the $H_{160}$-band image using the
SExtractor image analysis package (Bertin \& Arnouts 1996). A detection
threshold of $\mu_{H}=25.5$ mag arcsec$^2$ over 15 connected pixels was used. 
We
adopt MAG\_BEST from SExtractor as total magnitude of each detected object. Of
the sources extracted by SExtractor, objects located at the edge of WFPC2 
frames
and those, which we identified by eyes as noise peaks (mostly near the bright
objects) are rejected and removed from the final catalog.

 Figure 1 shows the raw number counts of $H_{160}$-band frame. From the figure,
we infer that the object detection is nearly complete to $H_{160}=$25.5-26.0. In
fact, the sensitivity of the frame varies over the field of view due to
variations in exposure time and NICMOS quantum efficiency (Dickinson et al.
2000). But this does not provide important effect in our analysis later on,
since we concentrate on the objects at least 1-1.5 mag brighter than above
completeness limit in this paper (see next subsection).
%
\begin{figure*}
  \begin{center}
    \epsfile{file=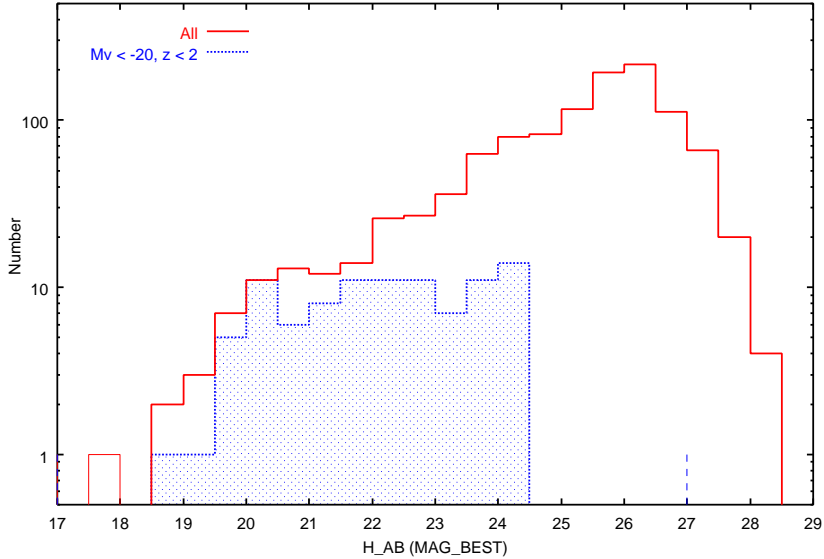,scale=0.6}
  \end{center}
  \caption{Raw $H_{110}$-band number counts of the Hubble Deep 
Field North (solid line).
Dotted line shows the galaxies with M$_{V}<-20$ at $z<2$.}\label{fig:NC}
\end{figure*}

 We cataloged the objects with $H_{160}<26$, and performed the photometry with
1.2$^{\prime\prime}$ diameter aperture for the all six $U_{300}$, $B_{450}$,
$V_{606}$, $I_{814}$, $J_{110}$, $H_{160}$-band images. The aperture position
for each object is fixed in order to measure the colors or broad band SED 
at the
same physical region of the galaxy. 631 objects were cataloged and their
magnitude was measured.

\subsection{Redshift determination \& Sample Selection}

 Cohen et al.(2000) compiled the HDF galaxies with the redshifts confirmed by
spectroscopy so far. We identified the corresponding objects in our $H_{160}<26$
catalogue, by comparing the coordinates of the objects in Cohen et al.'s
catalogue and ours. 139 out of 631 objects were identified and the 
spectroscopic
redshifts of Cohen et al. can be used. Cohen et al's sample was selected mainly
by standard $R \le 24$ mag, and limited to $z \ltsim 1.3$ except for the Lyman
Break Galaxies at $z>2$ (e.g, Lowenthal et al 1997, Dickinson 1998).

 For the objects with no spectroscopic redshift, we can estimate their 
redshifts
by using photometric redshift technique with the photometric data of the six
optical-NIR bands mentioned in previous subsection. We computed photometric
redshifts of those objects by using the public code of $hyperz$ (Bolzonella et
al. 2000). The photometric redshift is calculated by $\chi^{2}$ minimization,
comparing the observed magnitudes to the values expected from a set of model
Spectral Energy Distribution. The free parameters involved in the fitting are
the redshift, spectral type (star formation history), age, color excess (dust
extinction). We chose to use the SED model of Bruzual \& Charlot synthetic
library (GISSEL98, Bruzual \& Charlot 1993) and the  Calzetti extinction law
(Calzetti et al. 2000).  We excluded the stars that confirmed by spectroscopy
from our sample, and did not perform further star/galaxy discrimination for
fainter objects.

 For the objects with spectroscopic redshift, we also computed the photometric
redshifts in order to test the accuracy of photometric redshift estimate 
for our
filter set. Figure 2 shows the comparison between the spectroscopic redshifts
and the photometric redshifts for those objects in our $H_{160}<26$ catalogue.
%
\begin{figure*}
  \begin{center}
    \epsfile{file=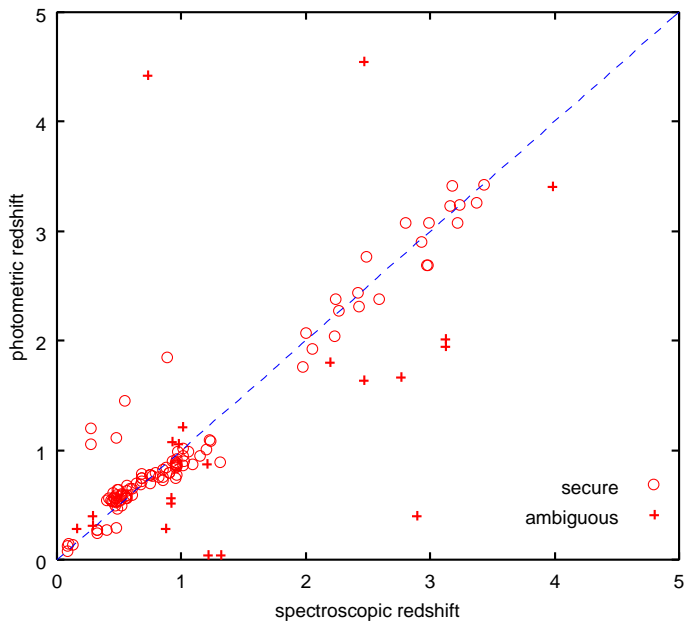,scale=0.5}
  \end{center}
  \caption{Comparison between the spectroscopic 
redshift and the photometric redshift
of the $H_{160}<26$ galaxies with spectroscopic redshift.
Circles represent the galaxies whose redshift likelihood function of
photometric redshift have single dominant peak, while crosses show those
sources for which photometric redshift technique failed to constrain their
redshift strongly.}\label{fig:specz}
\end{figure*}
%
 The spectroscopic and the photometric redshifts agree well within $\Delta z
= 0.11$ over the wide redshift range except for the several outliers in the
cases when the redshift probability function has a single dominant peak in the
photometric-redshift measurement.
 We note, however, that there may be some systematic offset at $z\sim1$;the
photometric redshift is tend to be smaller than the spectroscopic ones. 
Further, in Figure 3, we compared the $hyperz$ output with other photometric
redshift by Fernandez-Soto et al. (1999).
While our photometric redshift is calculated from HST $U_{300}$, $B_{450}$, 
$V_{606}$, $I_{814}$, $J_{110}$, $H_{160}$-band data, the
photometric redshift of Fernandez-Soto et al. (1999) is 
derived from HST $U_{300}$,
$B_{450}$, $V_{606}$, $I_{814}$-band data $+$ KPNO $J$, $H$, $K$-band data
(Dickinson 1998). 
Figure 3 shows the comparison between these two photometric redshifts
of $H_{160}<24.5$ galaxies which we can identified in common.
Although over all correspondence is relatively
good, there is the similar systematic offset at $z\sim1$. There are also more
than several outliers at $0 < z < 2$.
We will
study the possible influence of these features 
on our conclusion later in section 3.
\begin{figure*}
  \begin{center}
    \epsfile{file=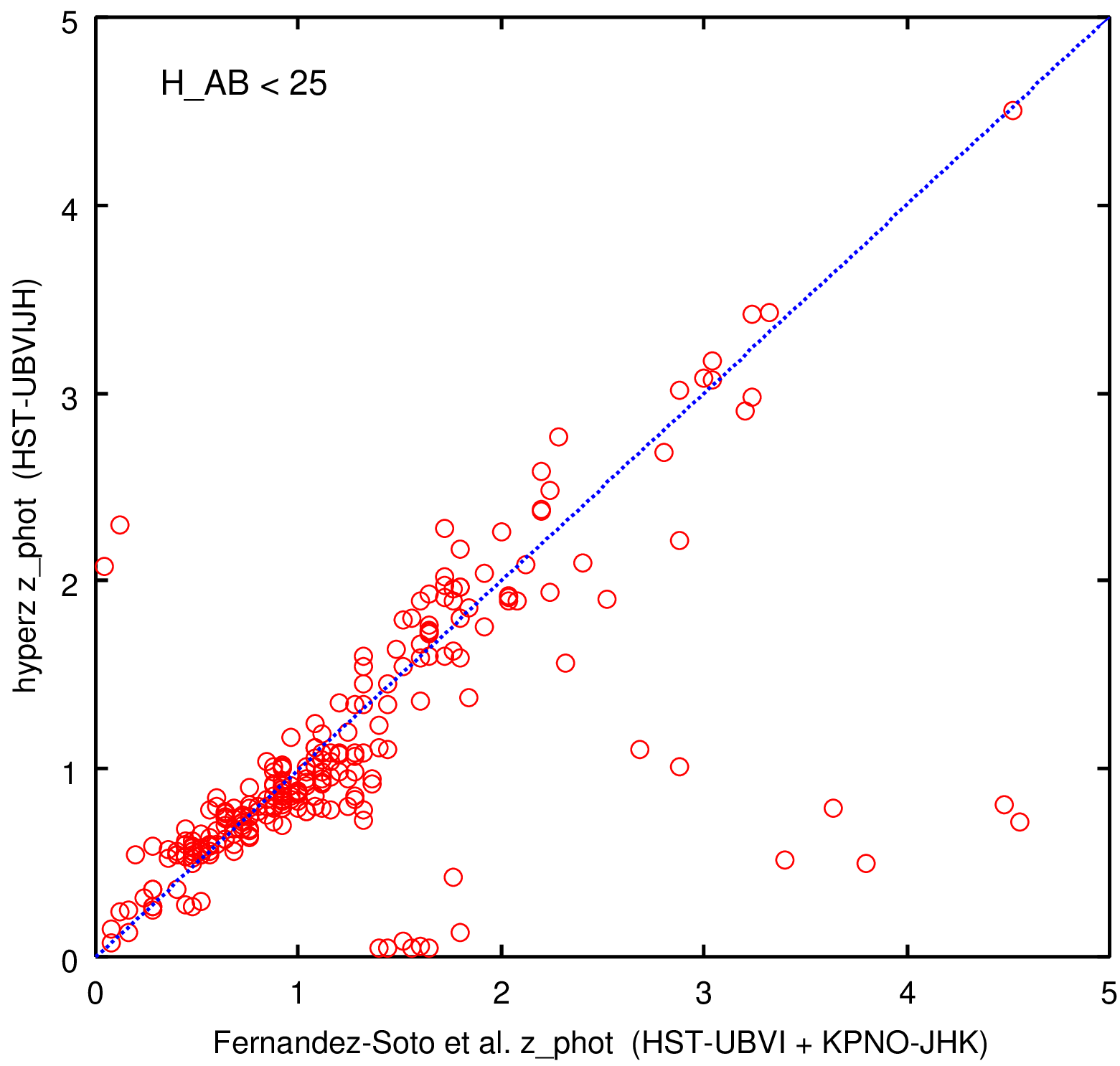,scale=0.5}
  \end{center}
  \caption{Comparison of different photometric 
redshifts of galaxies with $H_{160}<24.5$.
$hyperz$ output with HST WFPC2/NICMOS
$U_{300}$, $B_{450}$, $V_{606}$, $I_{814}$,
$J_{110}$, $H_{160}$-bands vs Fernandez-Soto et al.'s catalogue
with HST $U_{300}$, $B_{450}$, $V_{606}$, $I_{814}$-bands $+$ KPNO $J$, $H$,
$K$-bands.}\label{fig:photz}
\end{figure*}
%

 For the sample selection in studying the evolution of the Hubble sequence 
along the redshift, we adopted the rest-frame $V$-band magnitude, $M_{V}$. 
The $M_{V}$
values of the galaxies can be estimated from the observed data to $z=2$ by
interpolation and without extrapolation.
 Compared to the UV luminosity, rest-frame $V$-band luminosity is relatively
insensitive to the on-going star formation activity for a given stellar 
mass and the extinction by dust.
  $M_{V}$ (standard system) value of each object is calculated using the
redshift adopted above (spectroscopic or photometric), assuming H$_{0}=70$
km/s/Mpc, $\Omega_{\rm 0}=0.3$, $\Omega_{\Lambda}=0.7$.


 From the $H_{160}<26$ catalogue, 94 galaxies with $M_{V}<-20$ at $z<2$ were
selected as our final sample (50 with spectroscopic redshift, 44 with photometric redshift).
 In Figure 1, we also show the distribution of $H_{160}$-band magnitude of these
$z<2$, $M_{V}<-20$ galaxies. All the galaxies with $M_{V}<-20$ at $z<2$ are
brighter than $H_{160}=24.5$, which corresponds to $M_{V}\sim -20$ at $z=2$ for
our assumed cosmology, and thus at least 1-1.5 mag brighter than the 
turn around
in the number counts at $H_{160}=$ 25.5-26 mag. 
In Figure 4, we investigate the detection completeness of $H_{160}=24.5$
objects with various sizes 
by adding to the observed $H_{160}$-band frame
the artificial galaxies with de Vaucouleurs or exponential surface brightness
profile produced with $IRAF$ $ARTDATA$ package.
All these artificial galaxies are face-on to measure the completeness 
conservatively.
From the figure, our source detection seems to be complete to the half light 
radius of 0.8 arcsec for exponential profile, and 1.3 arcsec for de Vaucouleurs
profile, which correspond to 6.5 kpc and 10.5 kpc 
at $z\sim$1-2 respectively for our adopted
cosmology. Although some low surface brightness galaxies escape from our 
detection, 
the detection completeness of the galaxies in the Hubble sequence with the 
normal surface brightness must be very high above $H_{160}=24.5$.
\begin{figure*}
  \begin{center}
    \epsfile{file=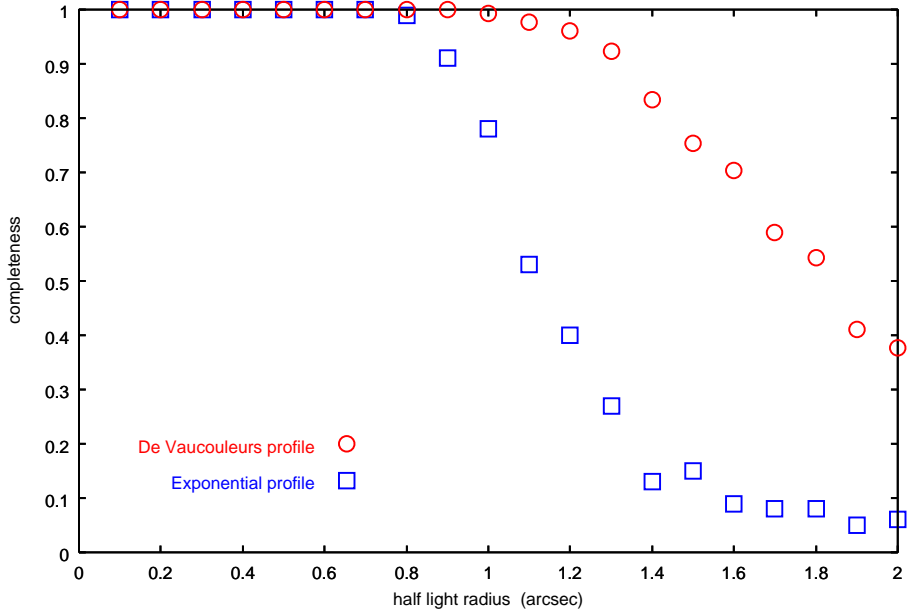,scale=0.65}
  \end{center}
  \caption{Detection completeness of $H_{160}=24.5$ galaxies as a function of 
half right radius. Circles show that for de Vaucouleurs surface brightness
profile, and squares represent that for exponential profile.}\label{fig:comple}
\end{figure*}
%

On the other hand, if the galaxy with $M_{V}<-20$
at lower redshift has $H_{160}>24.5$, the galaxy must have unrealistically
extremely blue SED between the observed band corresponding to rest-$V$ band 
and the
$H_{160}$-band, since the observed magnitude corresponding to rest-frame
$V$-band becomes relatively bright.
For example, we construct the constant SFR model spectrum of 1 Myr old with 
1/50 solar metallicity whose SED can be approximated as 
f$_{\nu}\sim \nu^{1.25}$, using GISSEL code, and calculate the expected 
H$_{160}$-band magnitude of the M$_{V}=-20$ galaxy with such extremely 
blue SED 
for each redshift. The result is showed in Figure 5, and it seems that 
even the galaxy with such a blue spectrum has H$_{160}<24.5$,
if it satisfies M$_{V}<-20$.
\begin{figure*}
  \begin{center}
    \epsfile{file=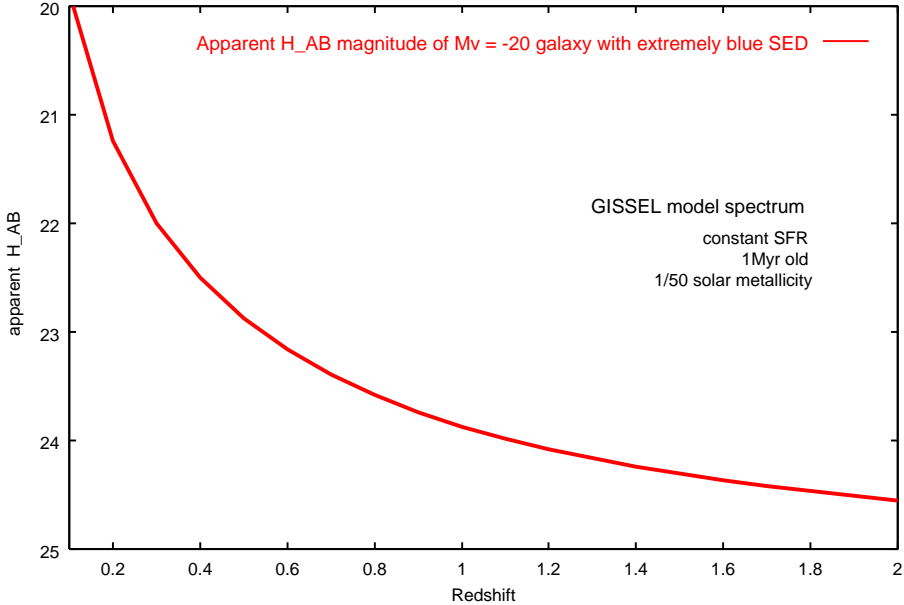,scale=0.65}
  \end{center}
  \caption{Predicted apparent $H_{160}$ magnitude of M$_{V}=-20$ galaxies with 
very blue SED as a function of redshift. The model spectrum used for 
calculation is constant SFR model of GISSEL synthetic library with 
1Myr age and 1/50 solar metallicity.}\label{fig:blue}
\end{figure*}
%


 Therefore, we believe that our $H_{160}<26$ source detection can completely
sample the $M_{V}<-20$ galaxies at $z<2$ in very strict sense (for the adopted
cosmological parameters).

\subsection{Morphological Classification}

 We could visually classify the galaxy morphology but it is rather difficult to
assign the morphological type to such faint objects and uncertainties can be
large and unknown systematic error could exist.
 Instead, we performed the quantitative morphological classification for the
$z<2$, $M_{V}<-20$ galaxies using the $I_{814}$, $J_{110}$, $H_{160}$-band HST
images so that the classification is made on the rest-frame $V$ to $I$-band
image for each object ($I_{814}$-band for $z<0.5$, $J_{814}$-band for
$0.5<z<1.0$, $H_{160}$-band for $1.0<z<2.0$), where the effects of
$K$-correction or star-forming region on their morphology are relatively small.
The classification method used is based on the central concentration index, $C$,
and the asymmetry index, $A$ (Abraham et al. 1996).

 Our procedure for measuring the $C$ and $A$ indices followed Abraham et
al.(1996), except that we adopted the centering algorithm of Conselice et
al.(2000). Early-type galaxies show strong central concentration in their
surface brightness distribution, while late-type galaxies have a lower
concentration. Those galaxies with a high asymmetry are classified as Irregular
galaxies.

 Since the $C$ and $A$ indices depend on object brightness, even if the
intrinsic light profiles are identical, we evaluated the values by comparing the
observed galaxy morphology with those of a set of simulated artificial ones that
have similar photometric parameters as each observed galaxy in the following
manner.

 First, a large number ($\sim$ 50000) of artificial galaxies with pure de
Vaucouleurs (bulge) or pure exponential (disk) profile with a wide range of half
light radius and effective surface brightness are constructed using the $IRAF$
$ARTDATA$ package and added to the observed images. In addition to magnitudes
and sizes (ISO\_AREA from SExtractor), the $C$ and $A$ indices of these
artificial galaxies were measured in an identical manner to the observed one.
 For each observed galaxy, the sample of the artificial galaxies with similar
isophotal magnitude, isophotal area and axial ratio  was constructed, and the
$C$ and $A$ indices of the observed object were compared with the distribution
of the indices of these artificial sample.

First, we examined the $A$ values of the representative bright irregular galaxies,
and adopted $A>0.12$ as the criterion of ``Irregular'' category. 
In fact, of the objects with $A>0.12$, 
those galaxies whose $A$ index is greater than
the median value of the artificial galaxy sample mentioned above at 2$\sigma$
level are classified as ``Irregular''. This criterion of 2$\sigma$ above the
sample of artificial galaxies with no intrinsic irregularity prevents relatively
faint galaxies with intrinsically symmetric light profiles from being classified
as irregular due to the effect of the background fluctuation.

 For the galaxies which are not entered into the category of irregular galaxy,
we then compare their $C$ indices with those of the artificial sample. Each
observed galaxy is classified as compact, bulge-dominated, intermediate,
disk-dominated, irregular, respectively, according to its $C$ value relative to
the de Vaucouleurs profile sample and the exponential profile sample.
 For this purpose,  the average value and the standard deviation of $C$ index of
the artificial galaxies are calculated for exponential sample and de Vaucouleurs
sample, respectively with $3\sigma$ clipping (hereafter, these values refer as
$C_{\rm dev}$, $\sigma_{\rm dev}$, $C_{\rm exp}$, $\sigma_{\rm exp}$).

 Table 1 shows the conditions for each morphological category.

 First, those galaxy whose $C$ index ranges within 1 $\sigma$ of {\it both} de
Vaucouleurs and exponential samples are classified as `cannot classify'. For
those sources, the both sample of artificial galaxies have similar (low) $C$
value due to their faintness, regardless of their surface brightness profile.

 For the other observed galaxies, 
the $C$ indices of the de Vaucouleurs and exponential
profile samples show distinctive distributions.
 If there are some observed galaxies whose $C$ value is significantly larger
than those of de Vaucouleurs profiles, they are to be classified as
`compact', there is no such an object in our observed $z<2$, 
$M_{V}<-20$ sample.
 On the other hand, we classify those galaxies that show significantly lower or
no concentration than that expected from exponential profile into the
`irregular' category. `Intermediate' galaxies are those whose $C$ value is
intermediate between  the distributions of de Vaucouleurs and exponential
samples, which are separate each other.
The galaxies whose $C$ values correspond to those of de Vaucouleurs sample are
classified as 'bulge-dominated', and those whose $C$ values locate within 
the range of exponential sample are entered into 'disk-dominated' category.

In Figure 6, we show the montage of those galaxies classified into each
category. The images in the figure are showed in pseudo-color 
produced from roughly rest-frame $U$, $B$, $V$-band
($U_{814}V_{606}I_{814}$-band for $z<0.5$, $B_{450}I_{814}J_{110}$-band 
for $0.5<z<1.0$, $V_{606}J_{110}H_{160}$-band for $1.0<z<2.0$).
\begin{figure*}
  \begin{center}
  \end{center}
  \caption{}
\end{figure*}
%
 The each row corresponds to each redshift bin, and in each row, galaxies are
aligned in order of the rest-frame $U-V$ color. As seen in Table 1, of 94
galaxies in $z<2$, $M_{V}<-20$ sample, 19 galaxies are classified as
bulge-dominated, 13 galaxies as intermediate, 32 galaxies as disk-dominated, 27
galaxies as irregular. The other 3 objects cannot be classified, and they are
all located at $z>1.5$.

 Figure 7 shows an example of the morphological classification procedure
mentioned above. In the figure, solid point shows the $C$ value of an observed
galaxy, while the histograms show the distributions of the de Vaucouleurs and
exponential artificial samples for this object. The dotted lines represent the
median $C$ value (vertical) and the standard deviation (horizontal) of each
artificial sample. The arrows
under the figure shows each morphological region for this observed
object derived from the conditions in Table 1, and  this object is classified 
as disk-dominated.
\begin{figure*}
  \begin{center}
    \epsfile{file=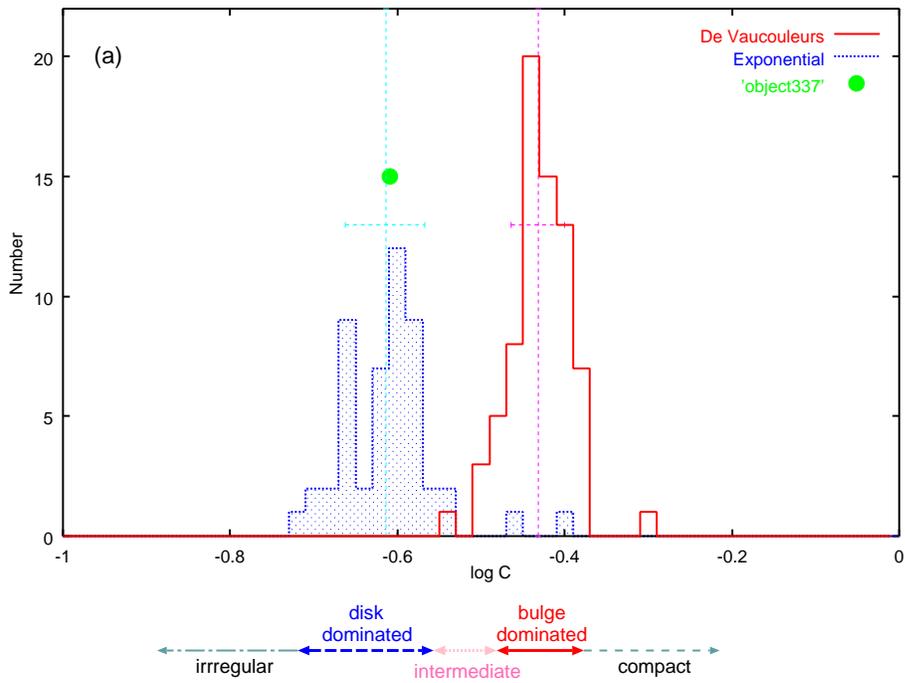,scale=0.65}
  \end{center}
  \caption{An example of morphological classification.
This observed object has low $A$ index value and is not irregular.
Solid point shows the $C$ index value of this object. Solid and dotted
histograms represent the distribution of the $C$ indices of the artificial
De Vaucouleurs and exponential samples. Dashed lines shows the average value
and the standard deviation of each artificial sample. Arrows under the 
figure represent the corresponding
morphological region of $C$ value derived 
from the condition in Table 1.}\label{fig:exam}
\end{figure*}
%

For Cohen et al.(2000)'s sample, van den Bergh et al.(2000) performed 
visual classification, and their result can be used for testing 
the relation between our $C$/$A$ classification and ordinal Hubble sequence
scheme.
This comparison is showed in Figure 8, where the morphological distribution 
of van den Bergh's classification is showed 
for each morphological category by $C$/$A$ classification in 
$I_{814}$-band according to van den Bergh et al.(2000).
Shaded regions represent the galaxies 
with peculiarity in van den Bergh classification, such as ``Sab pec''.
The figure shows that 
our quantitative $C$/$A$ classification correlates well with eyeball 
classification by van den Bergh, although some scatter exists.
Most `bulge-dominated' galaxies are classified as ``E'', `disk-dominated' 
galaxies corresponds to mainly spirals, and 
`irregular' galaxies are ``Irr'' or galaxies with
peculiarity in van den Bergh classification.
Intermediate galaxies scatter around the early-type spiral. 
\begin{figure*}
  \begin{center}
    \epsfile{file=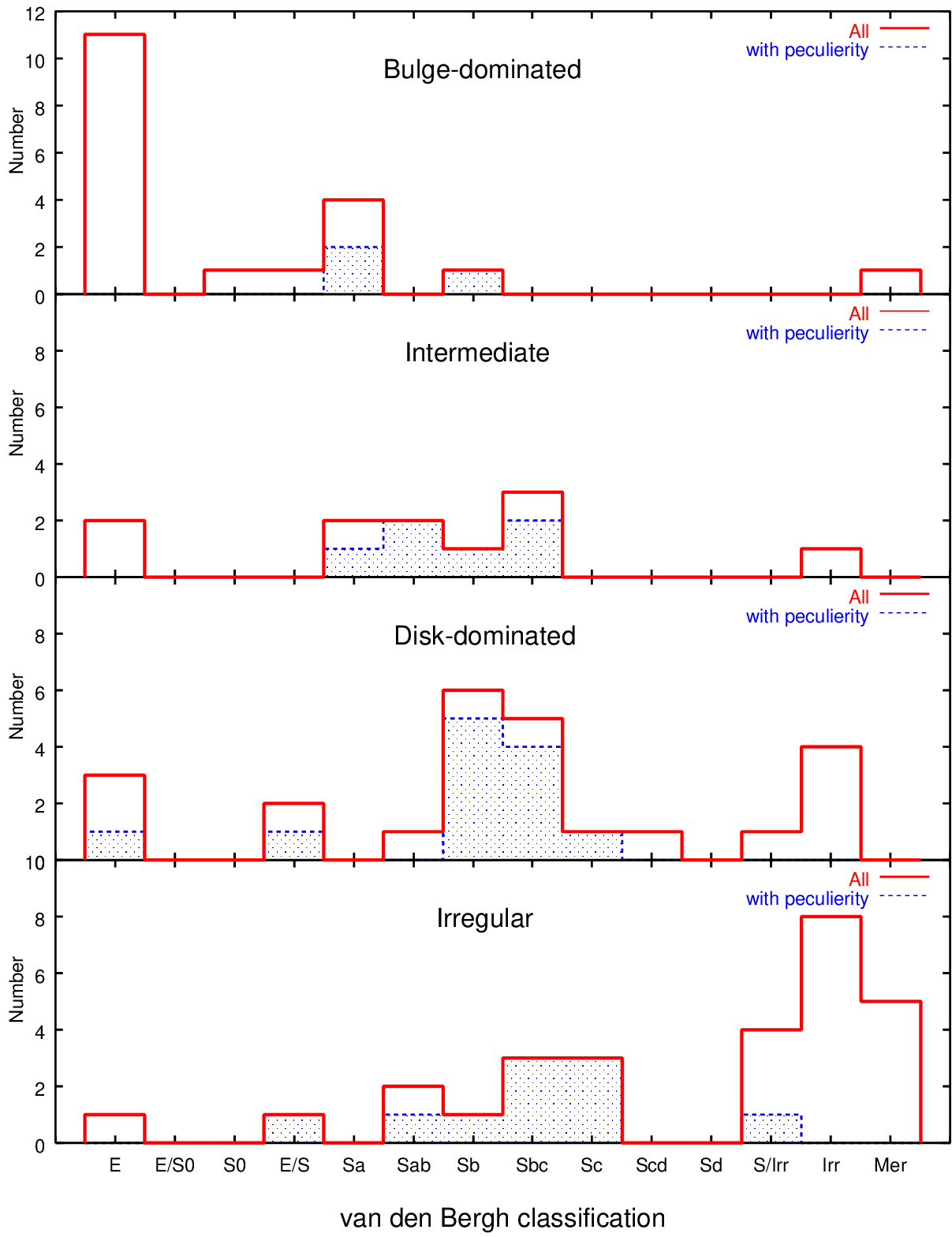,scale=0.65}
  \end{center}
  \caption{Comparison of morphological 
classification. Distribution of the visual
classification from van den Bergh et al. (2000) is showed for the galaxies of 
each $C$/$A$ classification morphological category. 
Shaded region represents the galaxies with peculiarity in van den Bergh 
classification.}\label{fig:vdB}
\end{figure*}
%

We prefer to use quantitative morphological classification, because 
it is not only more objective but also can be taken into account of 
the effect of magnitude/surface brightness bias for classification
by using the artificial galaxies with the same magnitude/surface brightness
as observed objects as mentioned above.

\section{Results}
 Using the volume-limited sample of $z<2$, $M_{V}<-20$ galaxies mentioned
above, we investigate the evolution of morphological 
comoving number density and
their rest-frame color distributions to $z=2$.

\subsection{Morphological Comoving Number Density}

 Figure 9 shows the evolution of the morphological comoving number density of
the $M_{V}<-20$ galaxies in the HDF-N. Each symbol represents the morphology
classified in the previous section, and error bars are based on the square root
of the observed number. We divided the redshift range $0<z<2$ into the four
$\Delta z=0.5$ bins, considering the number of the objects in each bin.
The cosmology of 
H$_{0}=70$ km/s/Mpc, $\Omega_{\rm 0}=0.3$, $\Omega_{\Lambda}=0.7$ is 
assumed.
\begin{figure*}
  \begin{center}
    \epsfile{file=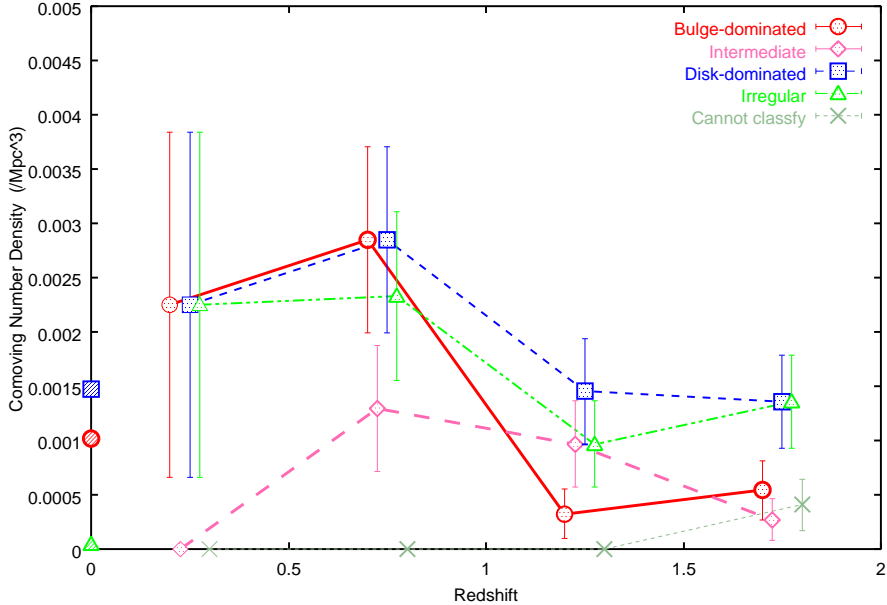,scale=0.65}
  \end{center}
  \caption{Evolution of morphological comoving number density
of galaxies with M$_{V}<-20$. Each symbol represent
morphological type. Circles: bulge-dominated, diamonds: intermediate,
squares: disk-dominated, triangles: irregular, crosses: cannot be
classified. Error bars are based on the square root of observed number.
Hatched symbols at $z=0$ show the number density
of local galaxies calculated from
the results of SSRS2 (Marzke et al. 1998): E/S0 (circles), Spiral (Squares),
Irr/Pec (triangles).}\label{fig:morNC}
\end{figure*}
%

 For comparison, the local number density of each morphology (E/S0, Spiral,
Irr/Pec) from Second Southern Sky Redshift Survey (Marzke et al. 1998) is
plotted at $z=0$, assuming the $B-V$ color of empirical spectral templates for
each morphology (Coleman et al. 1980).

 In Figure 9, except for the lowest redshift bin where the sample is very small
and statistical uncertainty is large, it is seen that the comoving density of
$M_{V}<-20$ galaxies decreases at $z>1$ in each morphological category. In
particular, the number density of bulge-dominated galaxies conspicuously
decreases to 1/5 between $0.5<z<1.0$ bin and $1.0<z<1.5$ bin.

In Figure 10, we show the evolution of the morphological {\it fraction} of 
M$_{V}<-20$ sample. As well as Figure 9, the local morphological fraction from
SSRS2 is plotted at $z=0$. Conspicuous features are that 
while at $0.5<z<1.0$, the fraction of irregular 
galaxies increases from local value, at $z>1$, that of bulge-dominated galaxies
decreases to about 1/3 of $0.5<z<1.0$ value. The fractions of irregular and 
disk-dominated galaxies slightly increase between $0.5<z<2.0$.
This evolution of the relative abundance of each morphological class seems to
reflect the differences of the luminosity or/and number evolution between
morphological classes.
\begin{figure*}
  \begin{center}
    \epsfile{file=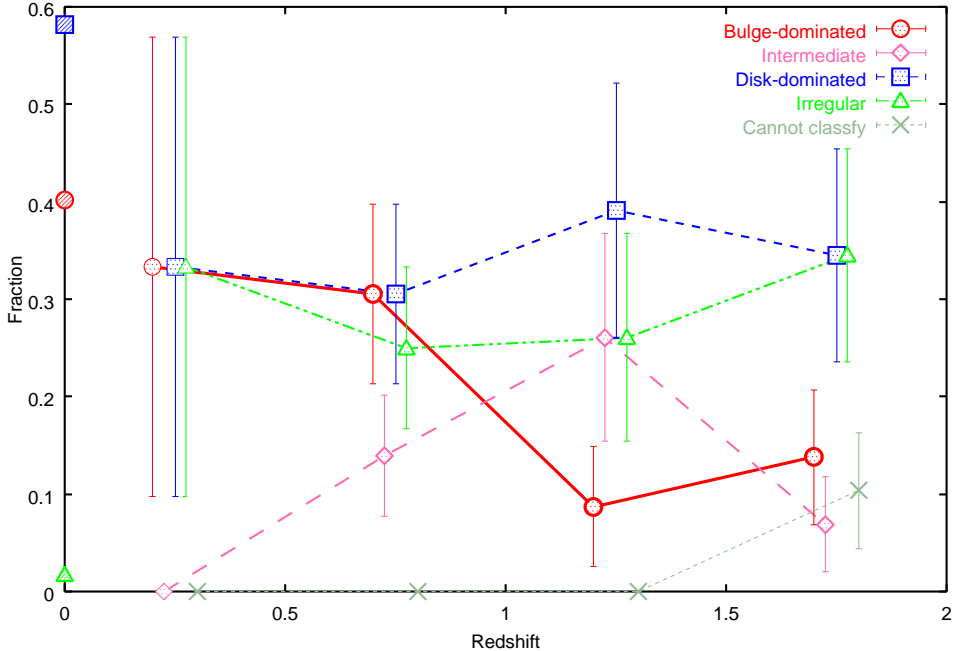,scale=0.7}
  \end{center}
  \caption{Evolution of morphological fraction of galaxies with M$_{V}<-20$.
Each symbol is the same as in Figure 9.
}\label{fig:morf}
\end{figure*}
%

 Before considering the origin of the decrease of number density of bright
galaxies, we check our results for some points.

 First, we investigated the behavior for the different set of the cosmological
parameters. For the Einstein de Sitter model, H$_{\rm 0}=50$, 
$\Omega_{\rm 0}=1.0$, 
the trend seen in Figure 9 does
not change significantly as shown in  Figure 11.  
\begin{figure*}
  \begin{center}
    \epsfile{file=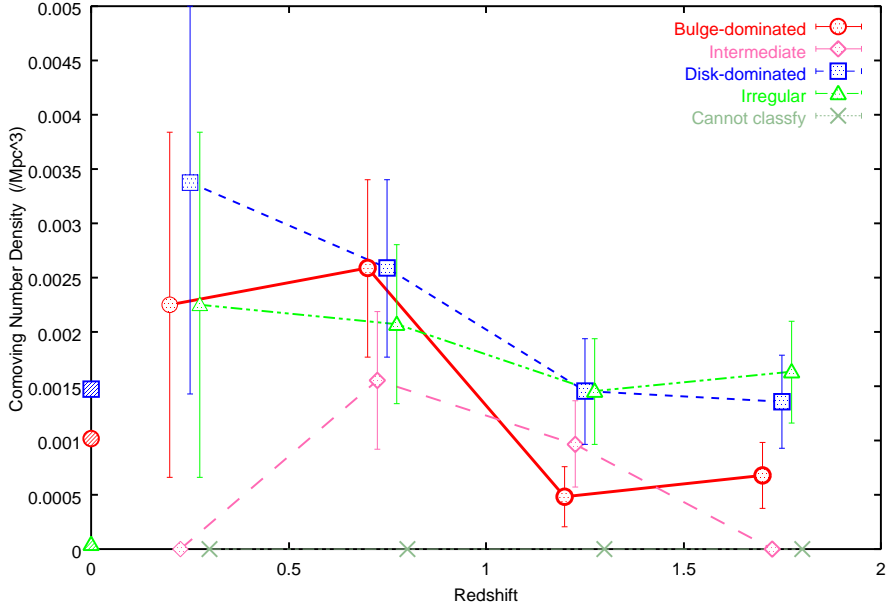,scale=0.65}
  \end{center}
  \caption{Same as Figure 9, but
for the cosmology of H$_0=$50, $\Omega_{\rm 0}=1.0$.
}\label{fig:morf}
\end{figure*}
%

 Second, as mentioned in section 2.4, we performed morphological classification
using different image for each redshift bin ($I_{814}$-band for $z<0.5$,
$J_{110}$-band for $0.5<z<1$, $H_{160}$-band for $1<z<2$), and this may cause
systematic effects on the evolution of morphological number density. 
In order to
check this point, we compared the classifications in the different bands.

Table 2 and 3 shows these comparisons, $I_{814}$-band classification vs.
$J_{110}$-band one for $0.3<z<0.8$ galaxies, $J_{110}$-band vs. 
$H_{160}$-band for $0.7<z<1.3$ galaxies.
Although there is some scatter, especially in Table 3 probably due to the
relative faintness of sample galaxies, the agreements in these figures are
relatively good and no strong systematic is seen. 
If we allow the excange between bulge-dominated and intermediate or 
disk-dominated and intermediate, 16/20 in Table 2 and 31/37 in Table 3 
agree with each other.
Therefore, the transfer of the
image used for classification at $z=0.5$ and $z=1$ does not seem to cause
systematic effect on each redshift bin.
%

 Third, as seen in Section 2.3, our photometric redshift seems to have
systematic offset relative to spectroscopic redshift and other
phtometric redshift at $z\sim1$, where the
decrease of the comoving number density of the bright galaxies is seen.
%
%
In order to investigate the effect of this feature, 
 we repeated the same analysis as in Figure 9 using the Fernandez-Soto et
al.'s photometric redshift. Of 166 $H_{160}<24.5$ objects with no spectroscopic
redshift, 151 objects can be found in Fernadez-Soto et al.'s catalogue. For the
other 15 objects, the photometric redshifts by $hyperz$ were used.
Of these 15 objects, 13 objects have $z_{phot}<2$, and three of the 13 objects
have M$_{V}<-20$. They are disk-dominated and irregular, and both located 
$1.5<z_{phot}<2.0$ range.

 The result is shown in Figure 12. The decrease of the 
number density in each morphology
 become slightly more marginal. 
While the number density of disk and irregular galaxies 
may be constant at $0.5<z<2$, 
the decrease in bulge-dominated galaxies seems to remain
significant in Figure 12.
\begin{figure*}
  \begin{center}
    \epsfile{file=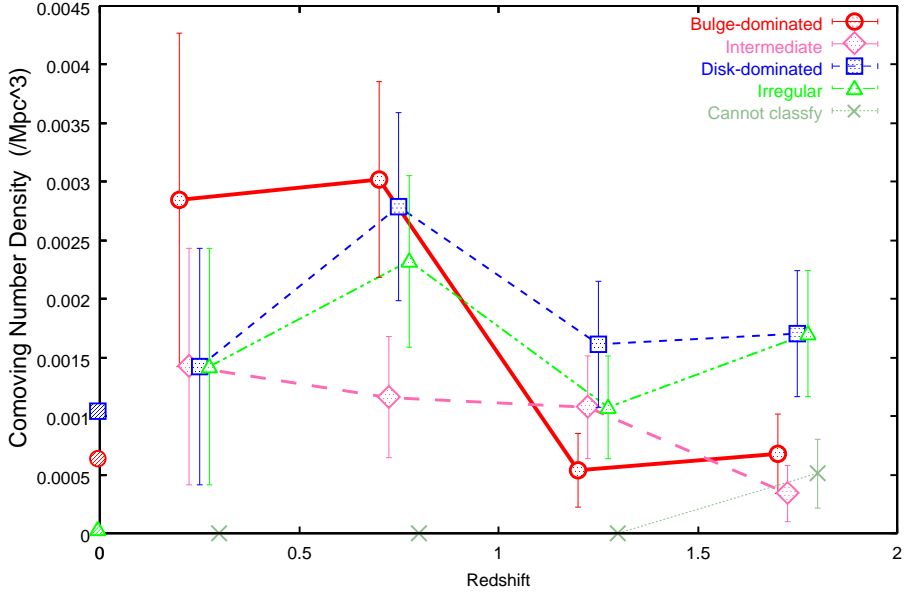,scale=0.65}
  \end{center}
  \caption{Same as Figure 9, but using the photometric 
redshifts from Fernandez-Soto
et al(1999) for the galaxies without spectroscopic redshift.
}\label{fig:morFer}
\end{figure*}
%
%

 The photometric redshift technique sometimes failed to constrain the redshift
of the objects especially for the blue flat SED and for the smoothly red SED
(produced by `blue flat SED $+$ dust extinction').
 In addition to the comparison between the different photometric redshifts with
different filter set mentioned above, we estimate the effect of this intrinsic
uncertainty of the photometric redshift by investigating the redshift 
likelihood function of each object outputted by $hyperz$.

 The redshift likelihood functions of the 166 $H_{160}<24.5$ galaxies with no
spectroscopic redshift were investigated, and the secondary peak whose 
amplitude
is grater than a fifth of the primary peak (i.e., best-fit $z_{phot}$) were
checked. Of 166 sources, 38 objects have these secondary peaks. For each
redshift bin in Figure 9 (and for each morphology), we counted possible number
of the objects which are escaped from or entered into each redshift bin if 
their
secondary peaks represent the true redshift.
The number of galaxies
in each redshift bin and these possible escaped and included number for the
secondary peak redshift are shown in Table 4.

 The effect of this uncertainty is negligible at $z<1$ due to the dominance of
spectroscopic sample in this redshift range. It is seen that at the uncertainty
of the disk-dominated and irregular galaxies is relatively large at $z>1.5$, 
while the bulge-dominated or intermediate galaxies is scarcely affected.

 In summary, we found that while the rapid decrease of the number density of 
the bulge-dominated galaxies with $M_V < -20$ at $z > 1$ may not be much 
affected by
the uncertainty of the photometric redshift, the decrease of the number-density
of the disk-dominated or irregular galaxies are rather marginal and to be 
widely tested and to be confirmed with the large sample in the future.
It is also very 
desirable to obtain the firm spectroscopic redshifts for the objects
at $1<z<2$.


\subsection{Rest-frame $U-V$ Colors}

 Galaxies in the Hubble sequence in the local universe also lie on the sequence
in the $U-V$ and $B-V$  two color diagram (Huchra 1977; Kennicutt 1983). In
other words, galaxy SED can be characterized by the rest-frame $U-V$ color
because the color is sensitive to the existence/absence of the largest features
in galaxy continuum spectra at near-UV to optical wavelength, namely, 4000 
\AA\  break or the Balmer discontinuity.
 If the stellar population in the galaxies changes along the redshift, the
Hubble 'color' sequence seen in the local universe must also change at high
redshift. In this subsection, we trace the color evolution of the galaxies with
$M_V < -20$ at $z < 2$.

  By using WFPC2 $U_{300}$, $B_{450}$, $V_{606}$, $I_{814}$ and NICMOS
$J_{110}$, $H_{160}$-band data, we can measure the rest-frame color blue ward
rest-frame $V$-band of the $z<2$, $M_{V}<-20$ galaxies without extrapolation.
 The rest $U-V$ color of each object is calculated from the best-fit SED
template in the photometric redshift technique and the standard Johnson $U$ and
$V$-band filter through put (for those with spectroscopic redshift, the similar
SED fit with broad band data was performed, fixing the redshift).
 Figure 13 shows the evolution of the rest $U-V$ color distribution (standard
Jhonson-like system) with M$_{V}$ value 
of the M$_{V}<-20$ galaxies for each morphology. For
comparison, at the top of each figure, 
the typical $U-V$ color of local galaxies is indicated for each morphology 
(Bershady et al. 2000).
\begin{figure*}
  \begin{center}
    \epsfile{file=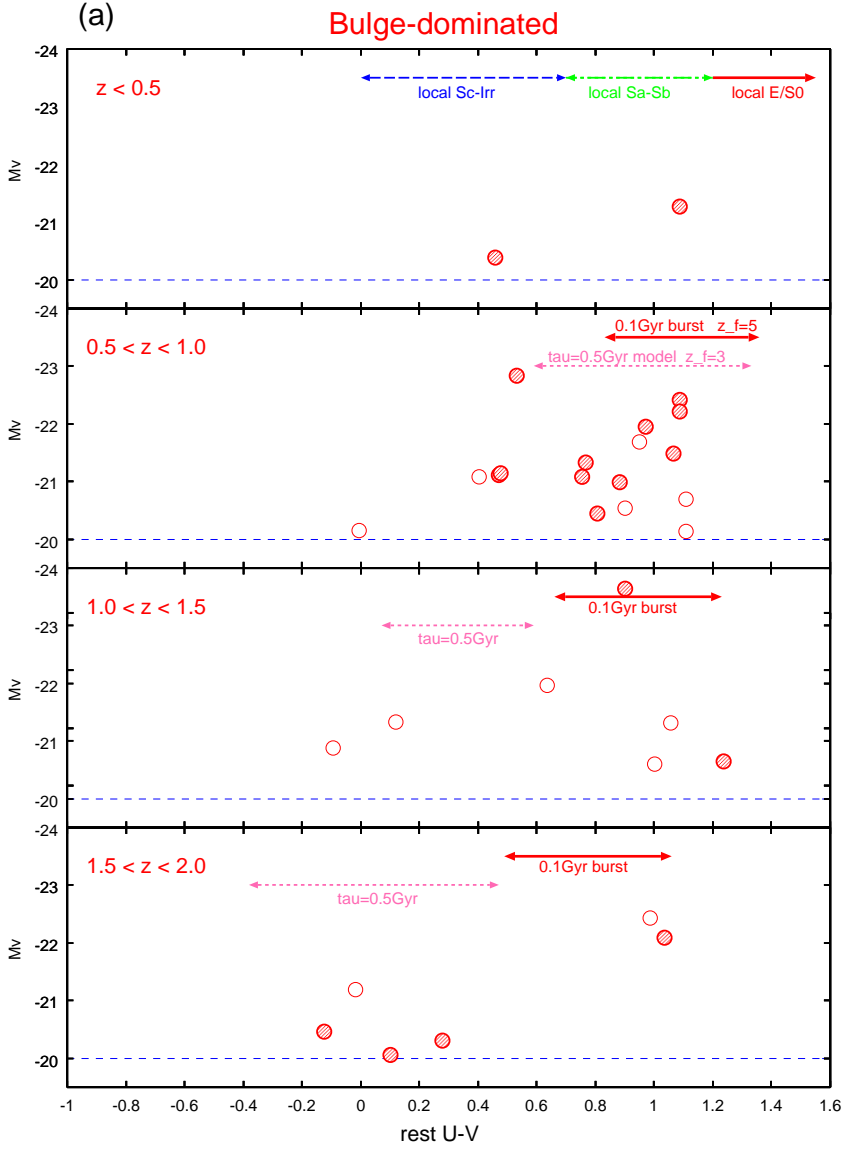,scale=0.6}
  \end{center}
  \caption{Rest-frame $U-V$ color distribution of M$_{V}<-20$ galaxies
in each morphology.
Each row corresponds to each redshift bin.
Vertical axis in each row represents the rest-frame $V$ absolute magnitude.
Arrows in the top row represent the corresponding range of
rest-frame $U-V$ color
of typical local galaxies with each morphological type.
}\label{fig:morFer}
\end{figure*}
\begin{figure*}[t]
  \begin{center}
    \epsfile{file=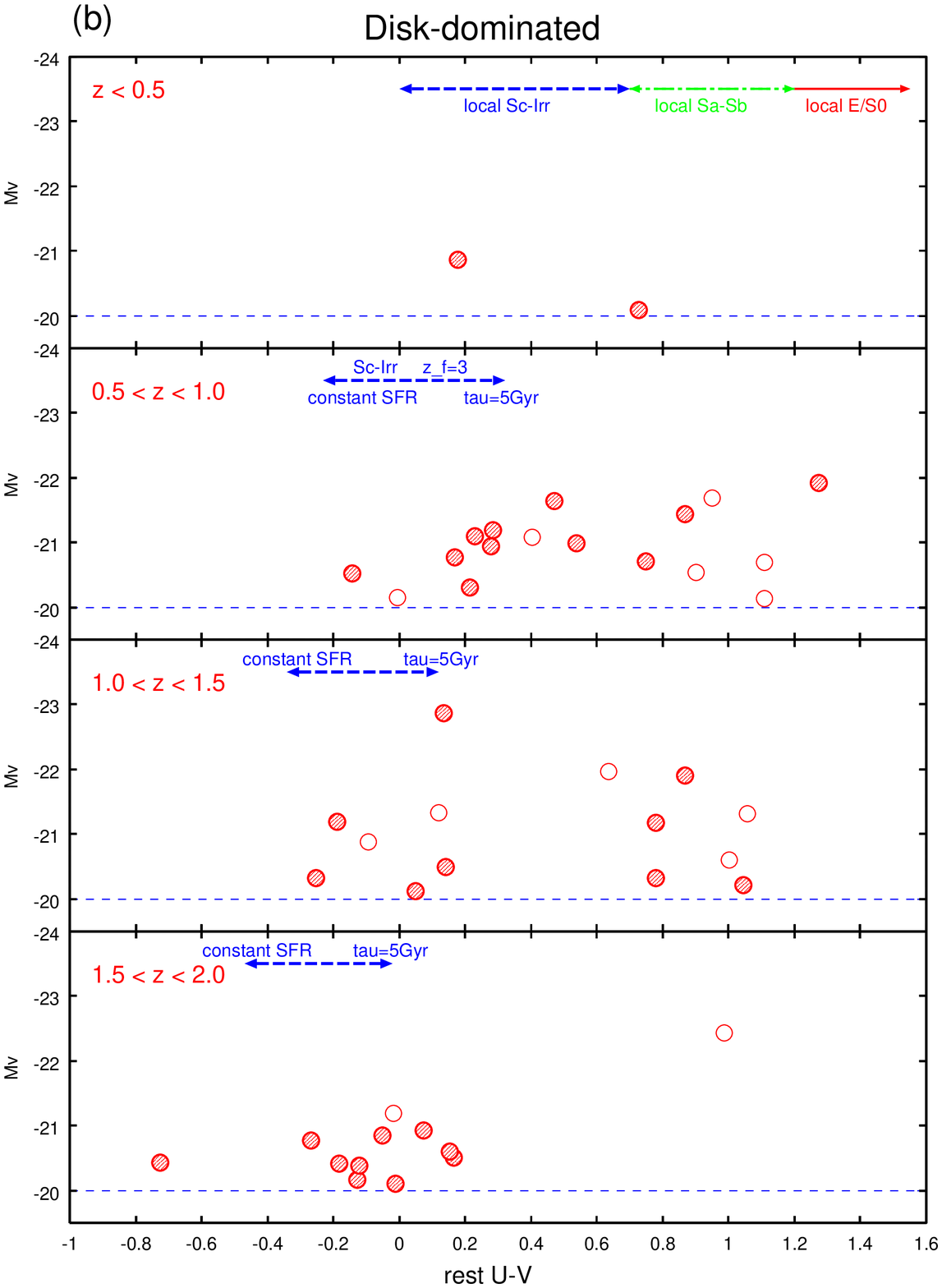,scale=0.6}
  \end{center}
  \begin{center}
  {\footnotesize Figure 13 continued}
  \end{center}
\end{figure*}
\begin{figure*}[t]
  \begin{center}
    \epsfile{file=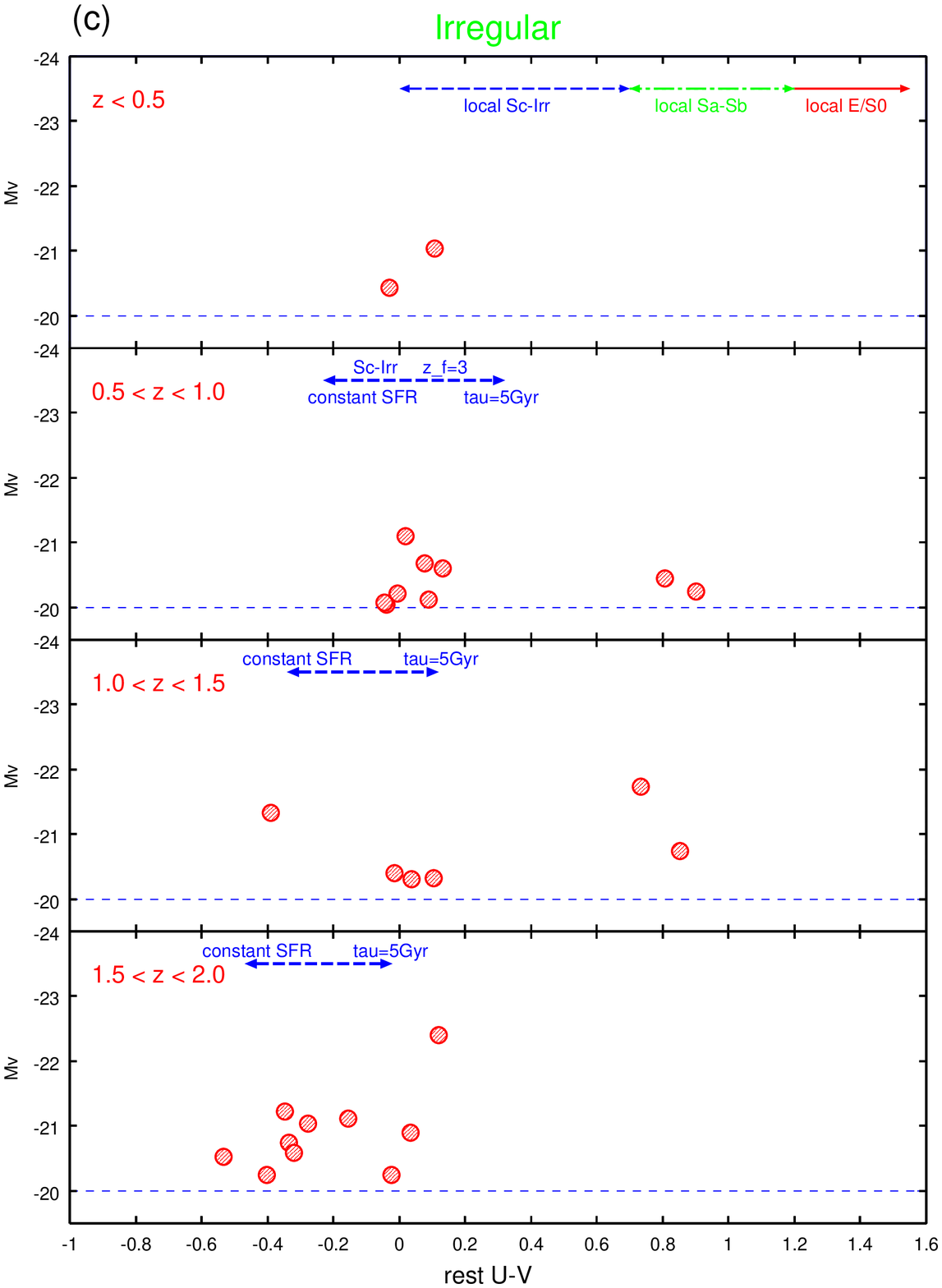,scale=0.6}
  \end{center}
  \begin{center}
  {\footnotesize Figure 13 continued}
  \end{center}
\end{figure*}
%

 In Figure 13a, the color distribution of bulge-dominated galaxies
(shaded circle) is showed with 
intermediate galaxies (open circle).
The bulge-dominated galaxies show the relatively broad color
distribution, from the similar red color with local early-type galaxies to blue
color of local Sc galaxies ($U-V\sim0.5$) at $z<1$. At $z>1$, although the
number of the galaxies in each bin is rather small, some blue ($U-V<0.3$)
bulge-dominated galaxies are seen, while there are also the galaxies with red
rest $U-V$ color which is consistent with passive evolution model ($U-V\sim1$)
with large formation redshift. For the intermediate-type galaxies,
the similar trend is seen.

 Figure 13b shows the similar diagram for disk-dominated galaxies (shaded 
circle) with intermediate ones (open circle), as Figure 13a. 
The rest $U-V$ color distribution of disk-dominated galaxies at
$z < 1.5$ cover the wide range which corresponds to local Sa-Irr color
($U-V\sim$0.1-1.2). There are also a few bluer galaxies with $U-V<0.1$, and
their fraction seems to increase with redshift.

 At $z>1.5$, all the disk-dominated galaxies in our sample 
have $U-V<0.3$. Since about 40\%
of the disk-dominated galaxies have $U-V>0.3$ at $z<1.5$, 
the significant change
of the rest $U-V$ color distribution seems to occur at $z\sim1.5$.

In Figure 13c, most irregular galaxies show the relatively blue color of $U-V
\sim 0$, which is similar to or slightly bluer than local Irr galaxies, while
there are also a few red irregular galaxies. 
At $z>1.5$, very blue galaxies with
$U-V<-0.2$ occupy significant fraction of irregular, and there is no irregular
galaxies with $U-V>0.3$ as well as disk-dominated galaxies.

 In Figure 14, in order to estimate the uncertainty of the rest-frame $U-V$
color due to the interpolation procedure, we compare the rest-frame $U-V$ color
derived from SED fitting mentioned above with that calculated from linear
interpolation between the nearest observed band data in f$_{\lambda}$.
\begin{figure*}
  \begin{center}
    \epsfile{file=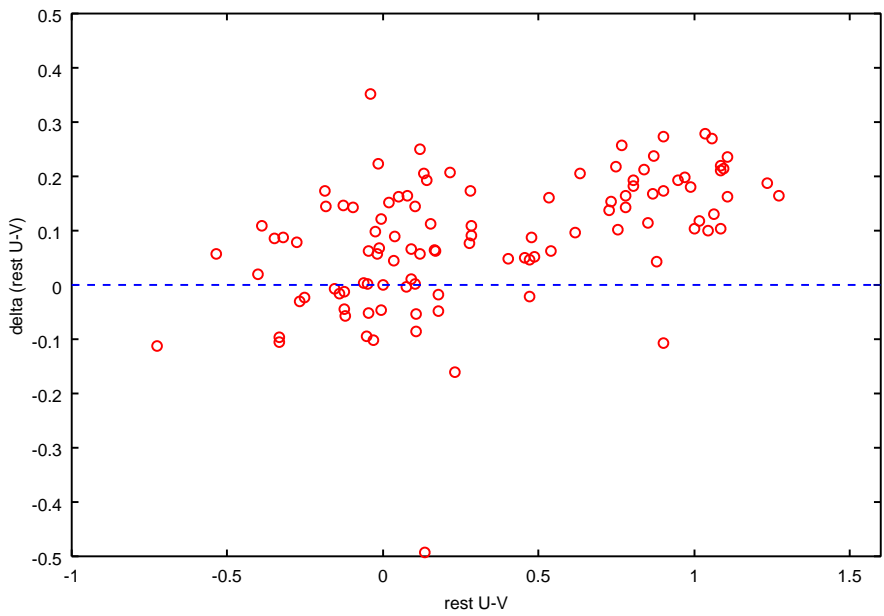,scale=0.65}
  \end{center}
  \caption{Differences between rest-frame $U-V$ colors derived from
SED fitting with six HST bands and the linear interpolation 
from the nearest observed bands to
rest $U$ and $V$-band, as a function of $U-V$ value (of SED fitting).
}\label{fig:morNC}
\end{figure*}
%
%
 Although there is some uncertainty due to the interpolation, 
$\Delta U-V$ is at most 0.2 mag (mean and standard deviation values are
 0.06 $\pm$ 0.13), and the trends seen 
in Figure 13 are not changed even if the colors derived from the linear 
interpolation of the nearest observed band data are used.

 Further, in order to evaluate the aperture effect, we repeated the photometry
with Kron aperture of $H_{160}$-band image of each objects, and compared
rest-frame $U-V$ color calculated from this result with those from the 1.2
arcsec diameter aperture photometry (Section 2.2) in Figure 15. It is seen that
the aperture effect does not affect strongly the results in Figure 13.
\begin{figure*}
  \begin{center}
    \epsfile{file=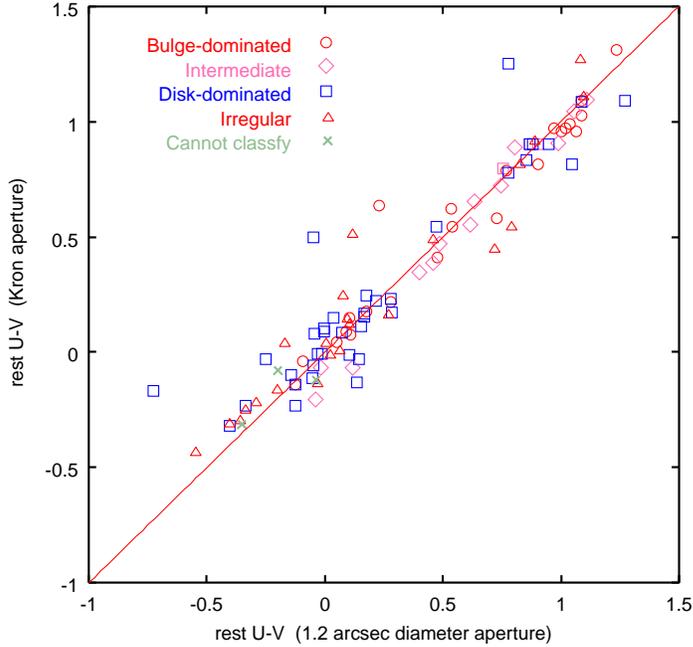,scale=0.5}
  \end{center}
  \caption{Comparison of rest-frame $U-V$ colors of M$_{V}<-20$, $z<2$ galaxies
measured with different aperture sizes:
1.2'' diameter aperture vs Kron aperture.
Each symbol represents morphological type.}\label{fig:morNC}
\end{figure*}
%
%
%
%
%
%

\section{Discussion}

\subsection{Bulge-dominated Galaxies}

 In Figure 9, the number density of the bulge-dominated galaxies decreases
rapidly at $z>1$. In fact, the number density at $z>1$ (two higher redshift
bins) is about two times smaller than the local number density of early-type
galaxies from the SSRS2 ($H_0$ = 70 km s$^{-1}$ Mpc$^{-1}$, $\Omega_{\rm 0}=0.3$, $\Omega_{\Lambda}=0.7$), although the
significance is two sigma level due to the small statistics. When compared with
the $0.5<z<1$ bin, the decrease become factor of five, which corresponds to
$\gtsim$3 sigma significance.

\subsubsection{Comparison with the Previous Studies at $z<1$}

 First, before considering the possible physical explanation of the result, we
compare our results of the earlier works although the galaxies treated in the
previous studies are 
brighter than those studied
in this paper at $z\gtsim0.8$.

 To $z=1$, Schade et al.(1999) investigated the evolution of the early-type
galaxies with $I_{\rm AB}<22$ 
in the CFRS/LDSS surveys selected by 2D-fitting morphological
classification with HST images, and found that the evolution of space density 
of
those galaxies is consistent with pure luminosity evolution (no number density
evolution) of $\Delta$M$_{B} \sim 1$ mag to $z=1$, which was also 
observed in their
size-luminosity relation and consistent with passive evolution model. 
The number
density of $0.5 < z < 1$ bin in Figure 9 is 2.5 times larger than the local
density of early-type galaxies in  SSRS2 survey in the same absolute magnitude
range, but considering the same luminosity evolution of $\Delta$M$_{B}\sim1$ to
$z=1$, it is also consistent with pure luminosity evolution.
In Figure 16a, we compare the evolution of the number density of 
bulge-dominated galaxies with the pure luminosity evolution model based on 
the local density of SSRS2 and the luminosity evolution of 
passive (single 0.1Gyr burst) model spectrum calculated by GISSEL code.
Although there may be some marginal excess, 
the number density of $0.5<z<1$ bin is consistent with 
the passive evolution model with high formation redshift.
For H$_{\rm 0}=50$, $\Omega_{\rm 0}=1.0$ cosmology (Figure 16b),
 the passive evolution model seems to underpredict 
the number density of early-type galaxies at $0.5 < z < 1$ 
(2.5$\sigma$ level). 
\begin{figure*}
  \begin{center}
     \epsfile{file=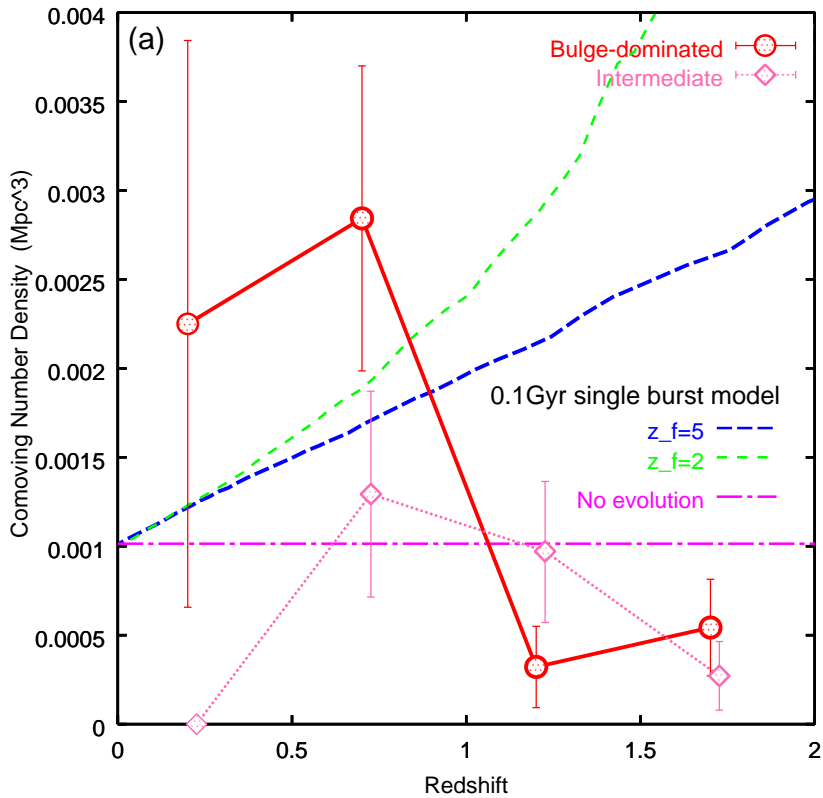,scale=0.6}
  \end{center}
  \caption{Comparison of the comoving number density evolution of 
bulge-dominated 
galaxies with pure luminosity passive 
evolution model. Circles represents number density of 
bulge-dominated galaxies, pentagon shows intermediate galaxies, which
are same as in Figure 9. Model 
predictions (dashed and dotted curves)
are calculated from local luminosity function of early-type 
galaxies from SSRS2 (Marzke et al. 1998) and the luminosity evolution 
derived from 0.1 Gyr single burst models of GISSEL library with formation 
redshift of two and five. a) for $H_0$ = 70 km s$^{-1}$ Mpc$^{-1}$, 
$\Omega_{\rm 0}=0.3$ 
$\Omega_{\Lambda}=0.7$ cosmology, b) for H$_0=$50, $\Omega_{\rm 0}=1.0$.}
\label{fig:morNC}
\end{figure*}
\begin{figure*}[h]
  \begin{center}
     \epsfile{file=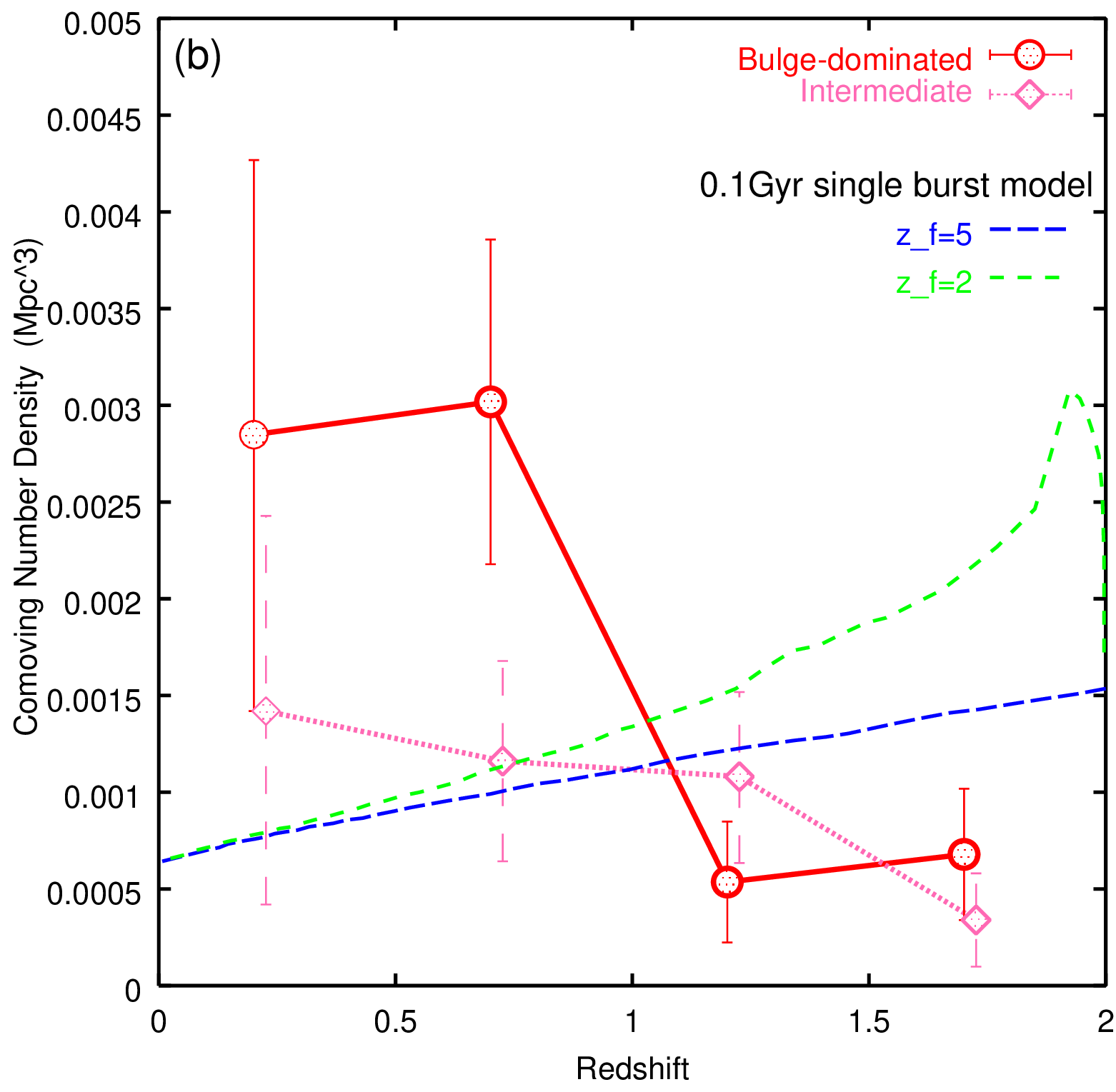,scale=0.55}
  \end{center}
   \begin{center}
    {\footnotesize Figure 16 continued}
   \end{center}
\end{figure*}
%
%

 Schade et al.(1999) also claimed that the significant fraction 
($\sim30$\%)
of their sample of early-type galaxies shows the rather bluer rest-frame $U-V$
colors than those expected from passive evolution model, and large
[O{\footnotesize II}] equivalent width. 
Although their selection of $I<22$ is about one magnitude brighter than 
our M$_{V}<-20$ selection at $z\gtsim0.75$,  
in Figure 13a, our bulge-dominated sample   
also shows the wide distribution of rest $U-V$ color, and some are
$\sim1$ mag bluer than those expected from  passive evolution model, which is
similar to Schade et al.

 Menanteau et al.(1999) investigated the optical-NIR color distribution of the
field early-type galaxies with $HK'\ltsim20$, 
which are selected by quantitative
classification with $C$ and $A$ indices in the HST images. They also found that
at $z<1$, the number density of luminous early-type galaxies is consistent with
that expected from the local luminosity function, but their optical-NIR 
color is
skewed blue-ward relative to the passive evolution model prediction. 
Their results agree with those of Schade et al. in that bright 
early-type galaxies is consistent with no number density evolution, 
but there are galaxies with bluer color than that expected from passive 
evolution model.


 At $z<1$, for low density totally flat universe, 
the result about the evolution of the bulge-dominated galaxies in 
our
sample seems to be consistent with those of other studies, although our HDF-N
data samples the fainter galaxies.

\subsubsection{Comparison with the Previous Studies at $z>1$}

 On the other hand, while the passive evolution model predicts that the 
galaxies
become more luminous as we approach the formation redshift, and the comoving
number density of the fixed absolute magnitude limited sample increase with
redshift, those of bulge-dominated galaxies decrease at $z>1$.
In Figure 16a, the deficit of bulge-dominated galaxies at $z>1$ relative to 
the model prediction calculated 
from local number density and passive luminosity evolution 
is significant.

 Several other studies investigated the number density evolution of early-type
galaxies at $z>1$ in the HDF-N.  Franceschini et al.(1998) studied the
ground-based $K$-selected ($K<20.15$) early-type galaxies selected by light
profile fitting morphological classification with WFPC2 image, and found that
their redshift distribution is suddenly decrease at $z\sim1.3$, which is
consistent with our result.  Rodighiero et al.(2001) recently added the same
$K<20.15$ sample in the HDF South WFPC2 and NICMOS fields to Franceschini et
al.' sample, and confirmed the similar results.  Franceschini et al.(1998)
claimed that this decrease of early-type galaxies is caused 
by the morphological
disturbance due to merging events and the dust obscuration in the starburst
phase of these galaxies.

 Note that their selection of $K<20.15$ is about 1 mag brighter at $z\sim1$, 2
mag at $z\sim2$ than our $M_{V}<-20$ selection. We thus confirm that the
decrease of number density of these galaxies at $z>1$ also seen for the less
luminous galaxies down to the absolute 
magnitude corresponding to $\sim$L$_{*}+1$.

 Zepf (1997) found that there is no extremely red ($V_{606}-K>7$) galaxies in
the HDF-N, and concluded that no passively evolving galaxy at $z>1$ with
formation redshift $z_{f}\sim5$ exists, and most early-type galaxies must have
significant star formation at $z<5$. In our volume-limited $M_{V}<-20$ sample,
there are a few galaxies at $z>1$ whose color is consistent with or 
slightly bluer 
than the passive evolution model with $z_{f}\sim5$ in Zepf (1997)
(corresponding rest $U-V\gtsim$1-1.2). Since the $V_{606}$-band traces
rest-frame $<3000$\AA\ at $z>1$,
if only small star formation activities are occurred in these galaxies, the
$V_{606}-K$ color can become bluer easily.

  Dickinson (2000b) compared the comoving number density of the $M_{V}<-19$
galaxies in the HDF-N between $0<z<1.37$ and $1.37<z<2$, using the same
WFPC2/NICMOS images as those used in this paper, and found the deficit of not
only the early-type galaxies but of all the bright galaxies at high redshift,
which is consistent with our result of Figure 9, although deficit of
disk-dominated and irregular galaxies is rather marginal if the uncertainty of
photometric redshift is considered in our analysis.
  He also pointed out that there are a few early-type galaxies with
$z_{phot}\sim1.8$ whose relatively red rest $B-V$ color is consistent with
passive evolution model with $z_{f}\sim4$ (in our sample, these galaxies
correspond to  bulge-dominated or intermediate galaxies with rest $U-V\sim 1.0$
in $z>1.5$ redshift bin), 
and that bluer early-type galaxies are also found, which is
seen in our Figure 13a, too.

\subsubsection{Morphological Number counts in NIR}

 In order to avoid the uncertainty of the photometric redshift, we also compare
the type-dependent number counts of the galaxies in the NICMOS $H_{160}$-band
image with the model prediction. Figure 17 shows the $H_{160}$-band 
number counts of the 
bulge-dominated and intermediate galaxies. The model prediction from 
local luminosity function of early-type galaxies (Marzke et al. 1998) and 
passive evolution (0.1 Gyr single burst with $z_{f}=5$) are also plotted.  
The upper curve represents the all early-type galaxies, and the middle and 
lower curves correspond to the contributions of those at $z<2$ and $z<1$, 
respectively. At $H_{160}>21.5$, where the contribution of $z>1$ galaxies 
increase in the passive evolution model, the observed number of bulge-dominated
galaxies is clearly deficient, although if $all$ intermediate galaxies are 
added, the observed number counts can explain the contribution of $z<2$ 
early-type galaxies. 
This result seems to 
suggest that the decrease of bulge-dominated galaxies at $z>1$ discussed 
above is not artifact due to the uncertainty of the photometric redshift 
technique.
\begin{figure*}
  \begin{center}
    \epsfile{file=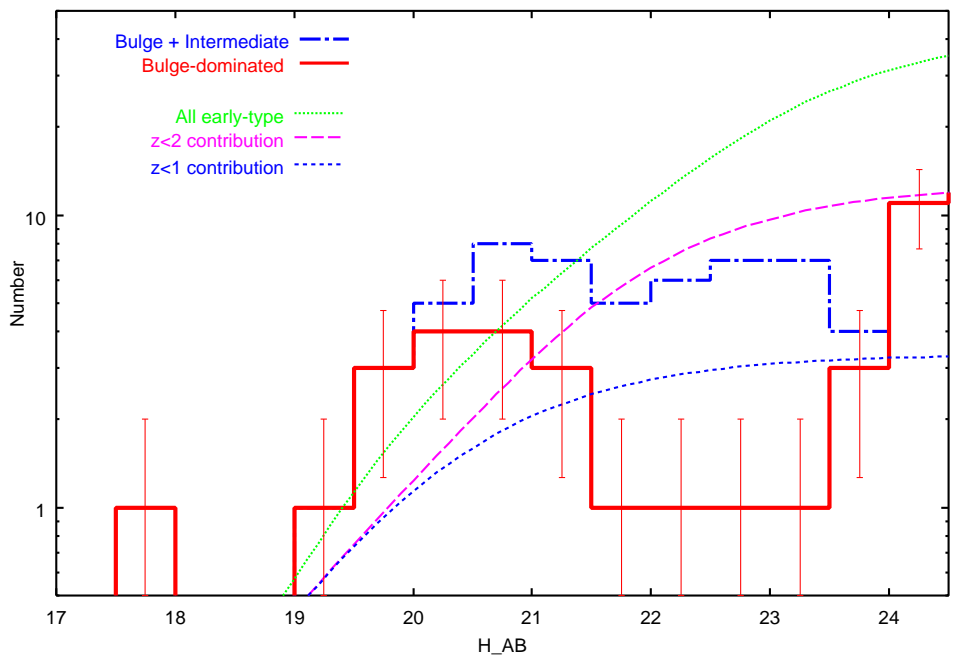,scale=0.7}
  \end{center}
  \caption{$H_{160}$-band number 
counts of bulge-dominated and intermediate galaxies.
Solid histogram shows bulge-dominated only, and dotted-dashed histogram 
represents bulge-dominated $+$ intermediate galaxies.
The curves show the prediction of the pure luminosity passive 
evolution model (see text).}
\label{fig:morNC}
\end{figure*}

\subsubsection{Implication}

 In summary, in the HDF-N, the number density of bright ($\gtsim$L$_{V}^{*}+1$)
early-type galaxies decreases at $z>1$, being contrary to the prediction of
passive evolution model with constant comoving number density. Further, their
wide color distribution
indicates the possibility that the mean stellar age of each field early-type
galaxy differs each other. It is seems that significant fraction of early-type
galaxies do not obey the passive evolution with high formation redshift, while
there are some galaxies whose color is consistent with those models.

 What causes the deficit of the red  bulge-dominated galaxies in the HDF-N at 
$z>1$ and how can we explain the blue colors of those {\it observed} 
at $z>0.5$ ?

 The decrease of number density of these galaxies at $z>1$ seen in Figure 9
indicates that significant fraction 
of these $\gtsim$L$_{*}$ early-type galaxies
does not seems to have the appearance of the representative bright early-type
galaxy at this redshift range.  Since the similarly bright irregular 
galaxies do
not show the corresponding rapid increase of the number density at $z>1$ in
Figure 9, and also not at the fainter galaxies in our $H_{160}<26$ catalogue
(e.g., M$_{V}<-19$ as we have confirmed although not shown in this paper in
detail), the morphological disturbance due to the simple merging scheme seems 
to be difficult to explain the observed decrease of early-type galaxies.

 As discussed in Franceschini et al.(1998), the dust obscuration may be
important at these epochs. However, if the deficit of the galaxies with M$_{V}
= -22$ at $z \sim 1.5$ is due to the dust extinction with $A_{V}\sim1$ mag 
for example, we may expect as many early-type galaxies with {\it red} 
colors due to the dust extinction. In our HDF-N sample, 
the observed number density for the 
galaxies $M_V < -20$ is still very small and they are tend to have {\it blue}
$U-V$ color. There may be some misclassification in our sample but not that the
colors of the galaxies in the rest of the sample with $M_V < -20$ (disk and
irregular) are even bluer than those of the early-type objects. 
Therefore, only the dust model
 does not seem to be compatible with the observed results.

 One possible explanation is that simply the formation (from the cold gas or
gas-rich small progenitors) epoch of the early-type galaxies is at $z\sim$
1.5-2 and they are formed in with relatively short time scale and with highly
spatially biased manner at higher redshift as expected for the CDM universe.

 The galaxies formed at $z>2$ may have red rest-frame $U-V$ color, but the
chance to observe such object may be low if they are strongly spatially 
biased.  
Note that
Steidel et al. (1998) and Adelberger et al. (1998) found the large 
clustering is
seen for the Lyman-Break galaxies at $z\sim3$. At $\sim10$ Mpc scale, the
1-sigma fluctuation of the galaxy density is 3 to 6 times the mass fluctuation
in the totally flat and the Einstein-de Sitter universe, respectively. 
Daddi et al. (2000) also found the very strong 
clustering for the galaxies with $K
\sim 19$ and $R-K > 5$ that are consistent with the passively-evolving galaxies
formed at $z>3$.

 If some galaxies are formed at rather low redshift, they may correspond to
the density peaks with lower peak height and the clustering is less strong than
those formed at higher redshift from higher density peak. They may appear in 
the absolute magnitude-limited sample but with relatively blue colors.
%

 Indeed, several bluer bulge-dominated galaxies whose rest $U-V$ colors are
similar to those of the local Scd galaxies (rest $U-V\sim0.6$) 
exist in our sample, 
and at $z>1$, there are still bluer
bulge-dominated galaxies.  We compared their color with single 0.1Gyr 
burst model of GISSEL code. 
In Figure 13a we show the prediction of the color range 
calculated from local color range with such passive evolution model 
with formation redshift of $z_{f}=5$.
Solar metallicity and Salpeter initial mass function are adopted.
 It is seen that the color of $U-V\ltsim0.6$ is clearly bluer than 
the passive model with $z_{f}=5$.
Although still lower metallicity may be considered, it is difficult that this
color of these bluer bulge-dominated galaxies is explained only by metallicity
variation since the local populations of early-type galaxies have solar or
sub-solar metallicity.
  It is plausible that these blue galaxies have some amount of star formation
activity or the younger mean stellar age than passive evolving galaxies.
In fact, at $z<1$, Schade et al (1999) detected strong [O{\footnotesize II}]
emission in the blue early-type galaxies.
In Figure 13a, 
we also plot the exponentially decaying SFR model with $\tau=0.5$Gyr,
and such model with same formation redshift can roughly explain the 
observed data.
Even in the cluster environments, a large fraction of the NIR-selected
galaxies show the bluer colors than that expected from 
passive evolution models, and the blue color 
seems to be due to some amount of the star-formation activity 
(Tanaka et al. 2000; Kajisawa et al. 2000; Haines et al. 2001; Nakata et al. 
2001).

 The blue bulge-dominated galaxies at $z>1$ tends to be relatively faint 
(M$_{V}\sim$-20) compared with $z<1$ bulge-dominated galaxies, although their 
number is very small. If these blue galaxies halt star formation and evolve 
passively soon after the observed epoch, these will become fainter and these 
are
not correspond to the ancestor of the $z<1$ bulge-dominated galaxies with
M$_{V}<-20$. If this is the case, the decrease of the number density of bright
bulge-dominated galaxies at $z>1$ can be more significant.

It should be noted that for H$_{\rm 0}=50$, $\Omega_{\rm 0}=1.0$ cosmology, 
the deficit of bulge-dominated galaxies relative to passive evolution model 
becomes to be rather marginal (Figure 16b). If all intermediate 
galaxies are added, 
the observed number density is consistent with the model prediction. 
In this case, as mentioned in Section 4.1.1, the observed number density 
significantly exceeds the model prediction at intermediate redshift. 
This may be explained as the cosmic variance due to 
the over density peak in redshift distribution 
discussed in Cohen et al.(2000). 


\subsection{Disk-dominated Galaxies}

   For disk-dominated galaxies, it is found that the rest-frame $U-V$ color
distribution significantly changes from $z \sim 0.5$ toward $z \sim 2$ 
and there
are some very blue galaxies with $U-V < -0.3$ while the comoving number density
does not show significant evolution between $0<z<2$. 
At $z> 1.5$, 
the colors of the galaxies are found to be very blue ($U-V < 0.3$), 
and few
objects have the $U-V$ colors such as seen for the local normal late-type
galaxies. It is very likely that we see the galaxies at very young stage at 
$z > 1.5$.

\subsubsection{Comparison with the Previous Studies at $z<1$}

 At $z<1$, our data suggest mild color and luminosity evolution for the
population, which is consistent with the previous results obtained by various
authors.

 Lilly et al.(1998) studied the disk-dominated galaxies selected by 2D-fitting
classification in the CFRS/LDSS survey, and found that the size function of the
relatively large (and bright) galaxies does not change at $0<z<1$ while the
rest-frame $B$-band luminosity at $z\sim0.7$ is $\sim$0.5 mag brighter than the
local value on the average. In Figure 18, we compare the 
comoving density of disk-dominated galaxies with such mild luminosity 
evolution (exponential decaying SFR model with $\tau=$7 or 5Gyr)
with constant number density.
At $z<1$, the observed number density is consistent with those model, as 
well as Lilly et al.
\begin{figure*}
  \begin{center}
    \epsfile{file=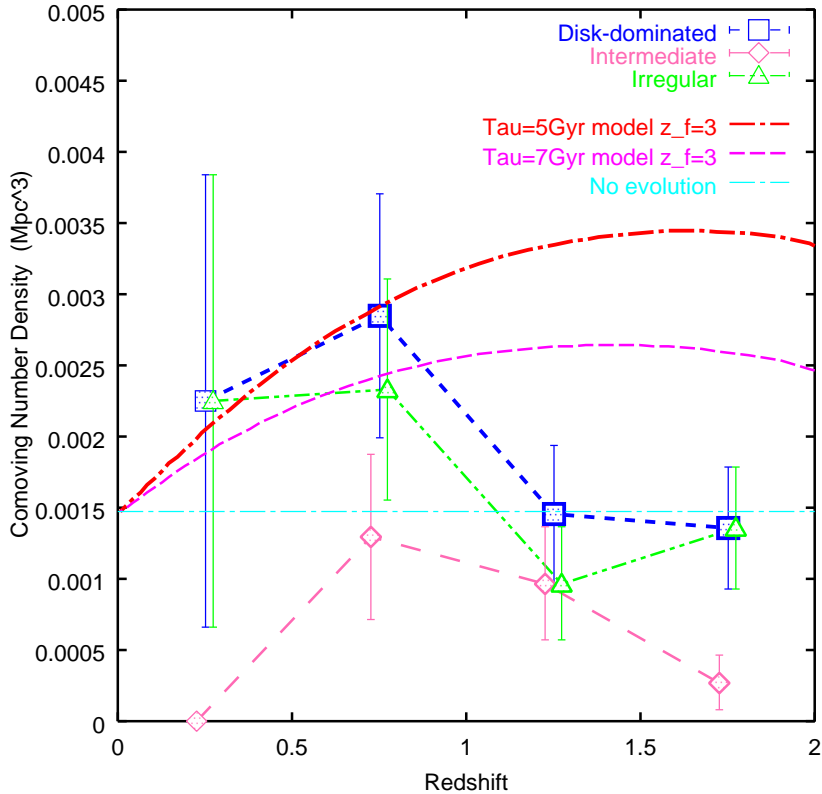,scale=0.6}
  \end{center}
  \caption{The similar figure with Figure 16 for disk-dominated galaxies.
Squares represent the comoving number density of disk-dominated galaxies,
pentagons represent that of intermediate galaxies, and triangles 
show irregular galaxies.
The curves show prediction of the pure luminosity evolution models 
derived from local luminosity function of spiral galaxies (from SSRS2) 
and the exponentially decaying SFR model with $\tau=$7 and 5Gyr.}
\label{fig:morNC}
\end{figure*}
%

 Vogt et al.(1996, 1997) confirmed that relatively bright ($\ltsim$M$_{*}+1$)
$z>0.5$ disk galaxies have the rotation curve similar to that of local disk
galaxies, and found that the similar rest $B$-band luminosity offset to local
sample of $\sim0.5$ mag in the Tully-Fisher relation and the surface brightness
evolution.
 These results seems to indicate that the bright disk galaxies in the present
universe have already formed at $z\sim1$.

 In Figure 13b, we show the rest-frame $U-V$ color of the disk-dominated
galaxies in the HDF-N. Lilly et al.(1998) pointed out that the average rest
$U-V$ color of the $z>0.5$ disk-dominated galaxies is slightly bluer (0.2-0.3
mag in their Figure 15) than local one.
 The average color of the disk-dominated galaxies in the $0.5<z<1$ bin in our
sample is $U-V\sim0.25$, which is slightly bluer than that obtained by Lilly et
al.(1998).  If we consider, however, the uncertainty due to the interpolation
and the aperture effects in evaluating the rest $U-V$ color (Figure 14 and 15),
the averages of the both samples is in fact consistent with each other.

 Lilly et al.(1998) interpreted the mild color evolution as the modest increase
of star formation in the bright disk-dominated galaxies at $z>0.5$, and pure
luminosity evolution model with the exponentially decaying star formation rate
with $\tau=$5Gyr and $z_{f}=4$ can roughly explain these observed trend.

\subsubsection{Evolution of Disk-dominated Galaxies at $z>1$}

At $z>1$, while the rest $U-V$ color distribution of the disk-dominated 
galaxies in the $1<z<1.5$ bin 
is wide-spread, which is consistent with mild evolution seen at $0.5<z<1$, 
$all$ disk-dominated galaxies in the $1.5<z<2.0$ bin have rest $U-V<0.3$,
which is clearly bluer than local late-type spirals.
In Figure 13b, we plot the model prediction of rest $U-V$ distribution
based on local value, whose upper limit is determined by $\tau=5$Gyr model 
and lower limit is determined by constant SFR model such that the model with
$z_{f}=3$ reproduces the local color range.
Even if formation redshift is replaced with $z_f=5$, predicted color range 
at $0.5<z<2.0$ shows little change.
Most disk-dominated galaxies at $z>1.5$ seem to be roughly consistent 
with the model. In the model such as $\tau=5$Gyr or constant SFR, 
these blue rest $U-V$ colors correspond to the age of $\ltsim2$Gyr.
Although these estimated age can easily change for different star formation
history or taking into account of dust extinction, the clear 
difference of rest $U-V$ color distribution
between $z<1.5$ and $z>1.5$ seems to reflect that 
the star formation activity of disk galaxies is stronger at $z>1.5$ 
and their stellar age is rather younger.

 Regarding the number density of late-type galaxies at $z>1$,  
Rodighiero et al.(2000) previously investigated the ground-based
$K$-selected ($K\gtsim20.5$) late-type galaxies in the HDF-N selected by light
profile fitting, and found the deficit of late-type galaxies at $z>1.4$.
 On the other hand, in our $M_{V}<-20$ sample, where the corresponding limiting
magnitude is $\sim$1.5 mag fainter than that in Rodighiero et al.(2000) at
$z\sim1.5$, the decrease of the comoving number density of disk-dominated
galaxies is only marginal, considering the uncertainty of the photometric
redshift technique (Section 3.1, Table 4).
 If we limit the sample to $M_{V}<-21.5$, however, we indeed see few
disk-dominated galaxies at $z>1.5$, and therefore the disk-dominated galaxies
in our sample have relatively faint (M$_{V}>-21$) rest $V$-band 
luminosity distribution at $z>1.5$ (Figure 13b).
The apparent inconsistency may be due to 
the luminosity-dependent evolution of the galaxies.

These bluer rest $U-V$ color and fainter $V$-band luminosity distributions of 
disk-dominated galaxies at $z>1.5$ suggest that 
we see these galaxies at very young stage near the formation epoch.

\subsubsection{Disk Galaxies with Red U-V Color}

 At $z<1.5$, Rodighiero et al.(2000) also pointed out that there are two kinds
of population, red and blue in their late-type galaxy sample. 
In our M$_{V}<-20$ sample, the relatively red disk-dominated galaxies 
exist at $z<1.5$, too.


The colors of rest $U-V\sim0.8$ of these red galaxies are significantly 
redder than the prediction of the $\tau=7$ or 5 Gyr model with $z_{f}=3$ or 5
mentioned above.
In these star formation history model, these colors correspond to 
$\sim$10 Gyr old at $0.5<z<1.5$, which is difficult to be considered in our 
assumed cosmology.
 Probably, their red $U-V$ colors are due to less on-going star formation
activity or stronger dust extinction effect. As mentioned in Rodighiero et
al.(2000), it is difficult to discriminate these two effects and determine the
mean age and the dust extinction, respectively with only UV-NIR photometry.

 In Figure 6c, 
we can see the montage of these red disk-dominated
galaxies. These galaxies show relatively 
red color all over the image, therefore if the
dust extinction causes their redness, the dust distributes over the whole
galaxy.

\subsection{Irregular galaxies}

  The relatively bright irregular galaxies in the HDF-N shows the similar
comoving number density (Figure 9) with disk-dominated galaxies at $z>0.5$,
which is much larger than that expected from local luminosity function. This
rapid increase of the irregular galaxies at $0.5<z<1.0$ is consistent with
several previous studies (e.g., Brinchmann et al. 1998, Abraham et al. 1996).

  The rest-frame $U-V$ color distribution of the irregular galaxies is also
similar to or slightly bluer than that of the disk-dominated galaxies, and at
$z>1.5$, all irregular galaxies have also $U-V<0.3$.

  Regarding the origin of irregular galaxies, Corbin et al.(2000) investigated
the same WFPC2/NICMOS images of the HDF-N as that used in this paper, and
suggested that most irregular galaxies are the minor mergers of 
disk galaxy with smaller companions. If this is real, 
the similarity of the evolution of the rest
$U-V$ color distribution of the disk-dominated galaxies and the irregular
galaxies may be expected assuming that 
the overall color distribution does not change significantly
during such a minor merger.
In Figure 18, if the sum of the disk-dominated and irregular galaxies is 
considered, the number density observed over $0.5<z<2$
can be explained by the pure luminosity 
evolution model of exponentially decaying SFR with $\tau=$7 or 5Gyr which 
based on the luminosity function of local disk galaxies.

 Walker et al.(1996) performed the N-body simulation of such mergers, and found
that as the result of these minor mergers, the bulge of the disk-dominated
galaxy is enlarged and the galaxy become slightly earlier-type spiral galaxy.
The red disk-dominated galaxies at $z<1.5$ may be the galaxies whose star
formation activity is suppressed as a 
result of gas consumption due to starburst
or/and gas stripping in these minor mergers.

Also, these red disk-dominated galaxies may be the progenitors of 
the red irregular galaxies at $z<1.5$. In this
scenario, the fact that at $z>1.5$, there is no red disk-dominated or irregular
galaxy may suggest that at $z>1.5$, there is no sufficient time when the
irregular galaxies due to the minor merger of (initial, pre-merger) blue
disk-dominated galaxies relax dynamically and return to (earlier and redder)
disk-dominated galaxies.
 In Walker et al's simulation, it takes about 1.5-2 Gyr since merger started
that the disk galaxy with stellar mass of 5.6 $\times$ 10$^{10}$ M$_{\solar}$ 
in minor
merger recovers the symmetric surface brightness distribution. If starbursts
occur in minor merger, the population formed in these bursts may have 
relatively red color, such a time after the bursts.

 On the other hand, at $z<1.0$, if these minor merger rate declines rapidly
forward $z=0$, the rapid decrease of the number density of irregular galaxies
may be able to be explained.
Further, if these minor merger occur in bulge-dominated galaxies at $z<1$, the
wide distribution of the rest $U-V$ color of 
these galaxies may also be expected
in this scenario.

\section{Conclusion: When the Hubble sequence appeared ?}
 
In order to understand the formation and evolution of the 
Hubble sequence, using the HST WFPC2/NICMOS archival data of the 
Hubble Deep Field North, we
constructed the volume-limited sample of $M_{V}<-20$ galaxies to $z=2$, and 
investigated the evolution of the distribution of 
morphology, luminosity, color of the M$_{V}<-20$ galaxies.

At $z<1$, as various studies have showed, the bright irregular galaxies 
increase 
rapidly from $z=0$ to intermediate redshift, while the observed number of 
the bright early-type and disk galaxies is consistent with mild luminosity 
evolution with constant number density to $z=1$.
Although their stellar population is younger, at least normal (non-irregular) 
galaxy portion of Hubble sequence have already formed at $z\sim1$ in the form 
such as seen in the present universe. 

At $1<z<1.5$, the number density of early-type galaxies decreases 
significantly, even if considering the uncertainty of the photometric redshift.
At least for $H_0$ = 70 km s$^{-1}$ Mpc$^{-1}$, $\Omega_{\rm 0}=0.3$ 
$\Omega_{\Lambda}=0.7$ cosmology, the observed number of early-type galaxies
at $z>1$ is deficit relative to the prediction of the pure luminosity evolution
model calculated from the local luminosity function.
Since that of disk-dominated and irregular galaxies does not 
show significant evolution, 
the overall number density of M$_{V}<-20$ galaxies decrease.
Although there is the uncertainty of the effect of the dust extinction on 
our sample selection, some of the M$_{V}<-20$ galaxies seem to have formed
at these epoch.
Similarly, the morphological fraction changes at these redshifts.
It indicates that the epoch of the formation of morphology is different
between the broad morphological categories, 
and most early-type galaxies with L$<$L$^{*}+1$ seem to form their appearance 
as E or S0 at later epoch than similarly bright disk galaxies.

On the other hands, at $1.5<z<2.0$, the distribution of rest-frame color 
of disk galaxies is rather bluer than that at $z=0$, and 
it suggests that these galaxies are very young. 
Since the number density evolution of 
these galaxies is unclear 
until the spectroscopic redshifts are obtained 
by the observation with large telescope, we cannot tell 
what fraction of disk galaxies forms at $1<z<2$, 
but their formation epoch do not seems to be so much earlier.

These strong evolution at $1.0<z<2.0$ suggests that Hubble sequence seen in
the present universe as $morphological$ $sequence$ is formed at these epoch,
although the formation epoch of each galaxy or its stellar population 
may be still earlier and wide-spread. 
Regarding the rest $U-V$ color, at $z>1.5$, Hubble sequence seems to be 
rather degenerate in blue range, 
although the marginal trend that the later-type
galaxies have bluer colors remains (Figure 19). 
Considering the HDF-N is at most 4 arcmin$^2$, these results are to be 
confirmed with much larger sample. 
\begin{figure*}
  \begin{center}
    \epsfile{file=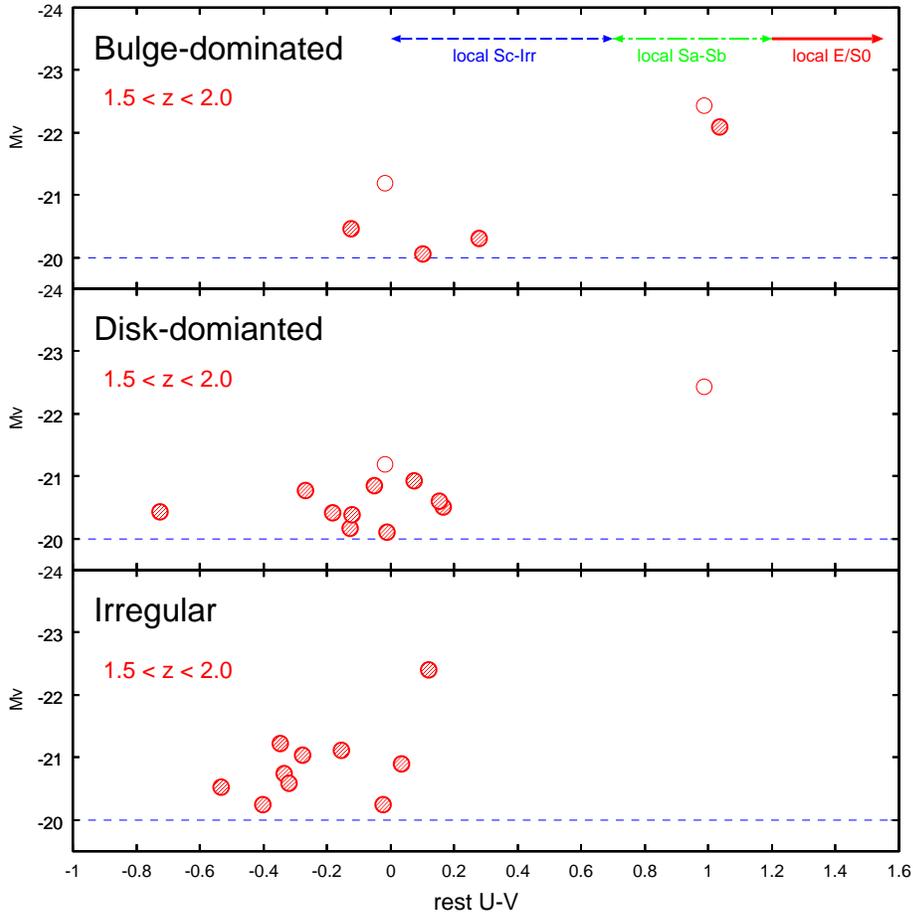,scale=0.65}
  \end{center}
  \caption{Similar with Figure 13 for the M$_{V}<-20$ galaxies 
at $1.5<z<2.0$ of 
each morphology. The shade symbols in each row show bulge-dominated, 
disk-dominated, irregular galaxies, respectively.
The open symbols in top and middle row represent intermediate ones.
}
\label{fig:morNC}
\end{figure*}


\newpage
\begin{table*}
   \caption{Condition of {\it C} and {\it A} indices for morphological classification}
\begin{center}
\begin{tabular}{ccc}
condition & morphology & observed number\\
\hline
\hline
$|C_{\rm obs}-C_{\rm dev}|<\sigma_{\rm dev}$ \& $|C_{\rm obs}-C_{\rm exp}|<\sigma_{\rm exp}$ & cannot classify & 3 \\
\hline
$C_{\rm obs} > C_{\rm dev}+3\sigma_{\rm dev}$ \& $C_{\rm obs} > C_{\rm exp}+\sigma_{\rm exp}$ & compact & 0\\
\hline
$C_{\rm dev}-\sigma_{\rm dev} < C_{\rm obs} < C_{\rm dev}+2\sigma_{\rm dev}$ & &\\
\& $C_{\rm obs} > C_{\rm exp}+\sigma_{\rm exp}$ & & \\
or & bulge-dominated & 19 \\
$C_{\rm exp}+\sigma_{\rm exp} < C_{\rm obs} < C_{\rm dev}-\sigma_{\rm dev}$\hspace{2cm} & &\\
\& $(C_{\rm obs}-C_{\rm exp})/\sigma_{\rm exp} > 2.5 \times (C_{\rm dev} - C_{\rm obs})/\sigma_{\rm dev}$ & &\\
\hline
$C_{\rm exp}+\sigma_{\rm exp} < C_{\rm obs} < C_{\rm dev}-\sigma_{\rm dev}$ & &
\\
\& $(C_{\rm obs}-C_{\rm exp})/\sigma_{\rm exp} > 0.4 \times (C_{\rm dev} - C_{\rm obs})/\sigma_{\rm dev}$\hspace{0.5cm} & intermediate & 13 \\
\& $(C_{\rm obs}-C_{\rm exp})/\sigma_{\rm exp} < 2.5 \times (C_{\rm dev} - C_{\rm obs})/\sigma_{\rm dev}$\hspace{0.5cm} & &\\
\hline
$C_{\rm exp}-2\sigma_{\rm dev} < C_{\rm obs} < C_{\rm exp}+\sigma_{\rm dev}$ & &\\
\& $C_{\rm obs} < C_{\rm dev}-\sigma_{\rm dev}$ & & \\
or & disk-dominated & 32 \\
$C_{\rm exp}+\sigma_{\rm exp} < C_{\rm obs} < C_{\rm dev}-\sigma_{\rm dev}$ & &\\
\& $(C_{\rm obs}-C_{\rm exp})/\sigma_{\rm exp} < 0.4 \times (C_{\rm dev} - C_{\rm obs})/\sigma_{\rm dev}$ & &\\
\hline
$C_{\rm obs} < C_{\rm exp}-3\sigma_{\rm exp}$ \& $C_{\rm obs} < C_{\rm dev}-\sigma_{\rm dev}$ & &\\
or & irregular & 27\\
$A_{\rm obs} > 0.12$ \& $A_{\rm obs} > A_{\rm art}+2\sigma_{\rm art}$ & &\\
\hline\\
\end{tabular}
\end{center}
\end{table*}

\begin{table*}
\caption{$I_{814}$ vs $J_{110}$-band classification for $0.3<z<0.8$}
\begin{center}
\begin{tabular}{p{4.5cm}ccccc}
\hline
\hline
&&$J_{110}$-band&&\\
$I_{814}$-band&bulge-dominated&intermediate&disk-dominated&irregular\\
\hline
bulge-dominated&4&2&2&0\\
intermediate&0&2&2&0\\
disk-dominated&1&0&2&0\\
irregular&1&0&0&4\\
\hline\\
\end{tabular}
\end{center}
\end{table*}
\ 
\vspace{15cm}
\ 
\begin{table*}
\caption{$J_{110}$ vs $H_{160}$-band classification for $0.7<z<1.3$}
\begin{center}
\begin{tabular}{p{4.5cm}ccccc}
\hline
\hline
&&$H_{160}$-band$$\\
$J_{110}$-band&bulge-dominated&intermediate&disk-dominated&irregular\\
\hline
bulge-dominated&6&3&2&0\\
intermediate&0&4&2&0\\
disk-dominated&0&3&5&0\\
irregular&1&1&2&8\\
\hline\\
\end{tabular}
\end{center}
\end{table*}
\begin{table*}
\caption{Number of galaxies in each redshift bin and possible escaped and included number for the secondary peak redshift}
\begin{center}
\begin{tabular}{p{1.9cm}ccccc}
redshift&bulge-dominated&intermediate&disk-dominated&irregular&cannot classify\\
\hline
\hline
0-0.5&4$\pm$0&2$\pm$0&2$^{\Large +1}_{\Large -0}$&2$\pm$0&0$\pm$0\\
0.5-1.0&13$\pm$0&5$^{+1}_{-0}$&12$^{+0}_{-1}$&11$\pm$0&0$\pm$0\\
1.0-1.5&3$\pm$0&6$^{+0}_{-1}$&9$^{+3}_{-1}$&5$\pm$0&0$^{+1}_{-0}$\\
1.5-2.0&4$^{+1}_{-2}$&2$\pm$0&10$^{+3}_{-9}$&10$^{+6}_{-6}$&3$^{+0}_{-2}$\\
\hline\\
\end{tabular}
\end{center}
\end{table*}

\small
\re
Abraham R. G.\ 1999, Ap\&SS, 269, 323
\re
Abraham R. G., van den Berg S., Glazebrook K., Ellis R. S.,
Santiago B. X., Surma P., Griffiths R. E.\ 1996, ApJS, 107, 1
\re
Adelberger K. L., Steidel C. C., Giavalisco M., Dickinson M., Pettini M., 
Kellogg M.\ 1998, ApJ, 505, 18
\re
Bershady M. A., Jangren A., Conselice C. J.\ 2000, AJ, 119, 2645
\re
Bertin E., Arnouts S.\ 1996, A\&AS, 117, 393
\re
Bingelli B., Sandage A., Tammann G.A.\ 1988, ARA\&A, 26, 509
\re
Bolzonella M., Miralles J.-M., Pell\'o R.\ 2000, A\&A, 363, 476
\re
Brinchmann J., Abraham G., Schade D., Tresse L., Ellis R. S., Lilly S., 
Le Fevre O., Glazebrook K., Hammer F., Colless M., Crampton D., Broadhurst T.
\ 1998, ApJ, 499, 112
\re
Bruzual A. G., Charlot S.\ 1993, ApJ, 405, 538
\re
Buta R.\ 1992a, in Morphological and Physical Classification of Galaxies, ed
 Longo G., Capaccioli M., Busarello, G.(Kluwer, Dordrecht) page1
\re
Buta R.\ 1992b, in Physics of Nearby Galaxies, ed Thuan T. X., Balkowski C., 
Van J. T. T.(Frontiers, Gif-sur-Yvette) page3
\re
Calzetti D., Armus L., Bohlin R. C., Kinney A. L., Koornneef J.,
Storchi-Bergmann T.\ 2000, ApJ, 533, 682
\re
Cohen J., Hogg D. W., Blandford R., Cowie L. L., Hu E., Songaila A.,
Shopbell P., Richberg K.\ 2000, ApJ, 538, 29
\re
Coleman G. D., Wu C.-C., Weedman D. W.\ 1980, ApJS, 43, 393
\re
Conselice C. J., Bershady M. A., Jangren A.\ 2000, ApJ, 529, 886
\re
Corbin M. R., Urban A., Stobie E., Thompson R. I., Schneider G.\ 2001, 
AJ, in press, {\it astro-ph}/0012192
\re
de Vaucouleurs G.\ 1959, in Handbuch der Physik, Vol. 53, ed Flugge, S.
(Springer-Verlag, Berlin) page275 
\re
Dickinson M.\ 1998, in The Hubble Deep Field, ed Lavio M., Fall S. M.,
Madu P.(Cambridge Univ. Press, Cambridge) page219
\re
Dickinson M.\ 2000a, Philosophical Transactions of 
the Royal Society of London, Series A, Vol. 358, no. 1772, page2001
\re
Dickinson M.\ 2000b, in Building Galaxies:From the Primordial Universe to the 
Present, ed Hammer F., Thuan T.X., Cayatte V., Guiderdoni B., Tranh Than Van
 J.(Ed. Frontieres, Paris) page257
\re
Dickinson M., Hanley C., Elston R., Eisenhardt P. R., Stanford S. A.,
Adelberget K. L., Shapley A., Steidel C. C., Papovich C., Szalay A. S.,
Bershady M. A., Conselice C. J., Ferguson H C., Fruchter A. S.\
2000, ApJ 531, 624
\re
Fernandez-Soto A., Lanzetta K. M., Yahil A.\ 1999, ApJ, 513, 34
\re
Franceschini A., Silva L., Fasano G., Granato L., Bressan A.,
Arnouts S., Danese L.\ 1998, ApJ, 506, 600
\re
Haines C. P., Clowes R. G., Campusano L. E., Adamson A. J.\ 2001, 
MNRAS in press, {\it astro-ph}/0011415
\re
Hubble E.\ 1926, ApJ, 64, 321
\re
Hubble E.\ 1936, The Realm of the Nebulae (Yale Univ. Press, New Haven)
\re
Huchra J. P.\ 1977, ApJ, 217, 928
\re
Kajisawa M. et al.\ 2000 PASJ, 52, 61
\re
Kennicutt R. C.\ 1983, ApJ, 272, 54
\re
Lilly S., Schade D., Ellis R., Le Fevre O., Brinchmann J.,
Tresse L., Abraham R., Hammer F., Crampton D., Colles M.,
Glazebrook K., Mallen-Ornelas G., Broadhurst T.\ 1998, ApJ, 500, 75
\re
Lowenthal J. D., Koo D. C., Guzman R., Gallego J., Phillips A. C.,
Faber S. M., Vogt N. P., Illingworth G. D., Gronwall C.\
1997, ApJ, 481, 673
\re
Marzke R. O., Geller M. J., Huchra, J. P.,  Corwin, H. G.\ 1994, AJ, 108,
437 
\re
Marzke R. O., da Costa L. N., Pellegrini P. S., Willmer C. N. A.,
Geller M. J.\ 1998, ApJ, 503, 617
\re
Menanteau F., Ellis R. S., Abraham R. G., Barger A. J., Cowie L. L.\ 1999,
MNRAS, 309, 208
\re
Nakata F., Kajisawa M., Yamada T., Kodama T., Shimasaku K., Tanaka I.\ 2001 
in preparation.
\re
O'Connell R. W.\ 1997, in The
Ultraviolet Universe at Low and High Redshift: Probing the Progress of
Galaxy Evolution, ed Waller H. W.(American Institute of Physics, New York)
 page11
\re
Oke J. B.\ 1974, ApJS, 27, 21
\re
Roberts M. S., Haynes M. P.\ 1994, ARA\&A, 32, 115
\re
Rodighiero G., Granato G. L., Franceschini A., Fasano G., Silva L.\ 2000,
A\&A in press, {\it astro-ph}/0010131
\re
Rodighiero G., Franceschini A., Fasano G.\ 2001, {\it astro-ph}/0101262
\re
Sandage A.\ 1975, in Stars and Stellar Systems, Vol.9, ed Sandage A., 
Sandage M., Kristian J.(Univ. Chicago Press, Chicago) page1
\re
Sandage A., Tammann G. A.\ 1987, A Revised Shapley-Ames Catalog of 
Bright Galaxies (Carnegie Insititute of Washington, Washingtion, D.C.)
\re
Schade D., Lilly S. J., Le Fevre O., Hammer F., Crampton D.\ 1996, ApJ, 464,
 79
\re
Schade D., Lilly S. J., Crampton D., Ellis R. S., Le F\'e vre O,
Hammer F., Brinchmann J., Abraham R., Colless M., Glazebrook K.,
Tresse L., Broadhurst T.\ 1999, ApJ, 525, 31
\re
Steidel C. C., Giavalisco M., Pettini M., Dickinson M., Adelberger K. L.\ 1996,
ApJ, 462, L17
\re
Steidel C. C., Adelberger K. L., Dickinson M., Giavalisco M., Pettini M., 
Kellogg M.\ 1998, ApJ, 492, 428
\re
Steidel C. C., Adelberger K. L., Giavalisco M., Dickinson M., Petteni M.\ 1999,
ApJ, 519, 1
\re
Tanaka I., Yamada T., Aragon-Salamanca A., Kodama T., Miyaji T., Ohta K., 
Arimoto N.\ 2000, ApJ, 528, 128
\re
van den Bergh S.\ 1997, AJ, 113, 2054
\re
van den Bergh S., Cohen J. G., Hogg D. W., Blandford R.\ 2000, AJ, 120, 2190 
\re
Vogt N. P., Forbes D. A., Phillips A. C., Gronwall C., Faber S. M.,
Illingworth G. D., Koo D. C.\ 1996, ApJ, 465, L15
\re
Vogt N. P., Phillips A. C., Faber S. M., Gallego J., Gronwall C.,
Guzman R., Illingworth G. D., Koo D. C., Lowenthal J.D.\
1997, ApJ, 479, L121
\re
Walker I. R., Mihos J. C., Hernquist L.\ 1996, ApJ, 460, 121
\re
Zepf S. E.\ 1997, Nature, 390, 377

\end{document}